\documentclass[twocolumn]{aastex631}
\usepackage{wrapfig, gensymb, amssymb}
\begin{document}

\title{Extensive Pollution of Uranus and Neptune's Atmospheres by Upsweep of Icy Material During the Nice Model Migration}
\correspondingauthor{Eva Zlimen}\email{ezlimen@ucsc.edu}
\author[0009-0006-9842-147X]{Eva Zlimen}
\affiliation{University of California, Santa Cruz\\ 1156 High Street\\ Santa Cruz, CA 95064, USA}
\author[0000-0002-4769-8253]{Elizabeth Bailey}
\affiliation{University of California, Santa Cruz\\ 1156 High Street\\ Santa Cruz, CA 95064, USA}
\author[0000-0001-5061-0462]{Ruth Murray-Clay}
\affiliation{University of California, Santa Cruz\\ 1156 High Street\\ Santa Cruz, CA 95064, USA}

\begin{abstract}
In the Nice model of solar system formation, Uranus and Neptune undergo an orbital upheaval, sweeping through a planetesimal disk. The region of the disk from which material is accreted by the ice giants during this phase of their evolution has not previously been identified. We perform direct N-body orbital simulations of the four giant planets to determine the amount and origin of solid accretion during this orbital upheaval. We find that the ice giants undergo an extreme bombardment event, with collision rates as much as $\sim$$3$ per hour assuming km-sized planetesimals, increasing the total planet mass by up to $\sim$$0.35\%$. In all cases, the initially outermost ice giant experiences the largest total enhancement. We determine that for some plausible planetesimal properties, the resulting atmospheric enrichment could potentially produce sufficient latent heat to alter the planetary cooling timescale according to existing models. Our findings suggest that substantial accretion during this phase of planetary evolution may have been sufficient to impact the atmospheric composition and thermal evolution of the ice giants, motivating future work on the fate of deposited solid material.
\end{abstract}

\section{Introduction}\label{sec:introduction}

The Nice model \citep{tsiganis2005origin,gomes2005origin, morbidelli2005chaotic} is a widely-invoked scenario for solar system formation, which was originally motivated by the long timescales required to form Uranus and Neptune at their current locations when gas-mediated accretion processes are not considered \citep[e.g.][]{helled2014formation}. While models of pebble accretion \citep[e.g.][]{lambrechts2012rapid} can grow the ice giants quickly \citep[e.g.,][]{frelikh2017formation}, Nice-model type upheaval scenarios remain popular due to their ability to produce dynamically excited small body populations in the solar system \citep[e.g.,][]{levison2008origin}.

A basic premise of the Nice model is that the four giant planets started in a much more compact configuration than seen today, with all four residing between $\sim$$5-20$ au. Jupiter and Saturn start out in a resonant configuration \citep[e.g.,][]{morbidelli2007dynamics}, which is disrupted by a chaotic upheaval due to interaction with a planetesimal disk extending from just beyond the orbit of the farthest planet to $\sim$$30$ au. Uranus and Neptune sweep through the disk, accreting and scattering planetesimals until dynamical friction brings them to rest at their approximate present-day positions. Numerical simulations \citep[e.g.][]{tsiganis2005origin} find that in $50\%$ of cases, Uranus and Neptune swap their initial orbital ranks. Whether Uranus and Neptune switched places is currently considered an open question. A scenario where Neptune started interior to Uranus' orbit could account for its greater mass, as the disk surface density is understood to decrease with radius \citep{helled2020interiors}.

During this sweeping outward migration proposed by the Nice model, Uranus and Neptune would have interacted with a substantially massive disk ($30-50$ $M_{\oplus}$) \citep{morbidelli2007dynamics}. While the proportion of material scattered versus accreted by these planets has been examined \citep{matter2009calculation}, the origin within the disk of these accreted planetesimals has not yet been considered in detail. To investigate the population of accreted planetesimals in more depth, we directly simulate collisions between test particles and planets. This is in contrast to \citet{matter2009calculation}, in which collisions were computed retroactively using the orbital elements of all particles at every time step to determine collision probabilities as described by \citet{wetherill1967collisions}. Additionally, we focus on late-stage accretion, with the inner edge of our planetesimal disk at a greater radial distance, thus excluding some planetesimals which may already have been incorporated into the planets at earlier stages of formation.

In this work, we derive the extent of late-stage accretion during a Nice model orbital migration using direct $N$-body simulations. We investigate whether this ``late veneer" of planetesimal accretion could have contributed substantial heavy element pollution to the atmosphere and/or envelope of the ice giants. We determine the relative differences in accretion expected between the two ice giants in different migration scenarios, particularly the two cases where Uranus and Neptune either swap or maintain their initial order.   

The ice giants Uranus and Neptune have broadly similar physical properties, such as radius, mass, and mean density. Both planets’ atmospheres are dominated by hydrogen, helium, and methane \citep{Guillot_2015}. While a protosolar abundance of helium relative to hydrogen is consistent with observations of both planets \citep{Conrath1991helium,Conrath1987helium}, their atmospheres are enriched in methane by over $50\times$ compared to the protosolar value \citep{baines1990atmospheric, sromovsky2011methane, karkoschka2011haze}. 

Moreover, similar interior structures are generally predicted for these planets: an exterior H$_{2}$ dominated envelope encasing a heavy-element enriched deep interior and, in some cases, a separate rocky core \citep{podolak1991models, hubbard1995interior, nettelmann2013new, bailey2021thermodynamically}. This comparison of interior models to the gravitational fields of these planets \citep{jacobson2014orbits, tyler1989voyager,lindal1992atmosphere} robustly demonstrates that the envelope of Neptune is significantly enriched in heavy elements compared to the envelope of Uranus. Due to the intermediate density of these planets, there is not a unique composition profile that satisfies the measured gravity fields \citep[e.g.][]{podolak1991uranus, de2013constraining}. For the sake of clarity, it is crucial to note that the ``atmosphere'' is generally defined  as a relatively thin layer exterior to the ``envelope,'' which extends deep into the interior.

In contrast to the dipole-dominated fields of other solar system dynamos, the relatively quadrupole-dominant fields of Uranus and Neptune  \citep{connerney1987magnetic, connerney1991magnetic} appear to suggest the two ice giants have distinctly similar interior convection patterns, although various different scenarios can be implicated in producing this type of field geometry. A convecting thin shell atop a stably stratified interior \citep{stanley2004convective, stanley2006numerical} has been a widely considered scenario. Alternatively, turbulent models \citep{soderlund2013turbulent} have also been suggested. Particularly given the distinctly similar magnetic fields of Uranus and Neptune, it is widely considered to be a paradox that the observed heat fluxes of these planets are vastly different. Both Voyager 2 \citep{conrath1989infrared, conrath1991thermal, pearl1991albedo} and ground-based \citep{loewenstein1977effective, loewenstein1977far} measurements have revealed that while Neptune has a significant heat flux, the heat flux from Uranus is much lower, and could be consistent with zero given current data, though an analysis which takes into account the wavelength dependence of Uranus' Bond albedo, influencing the  radiant energy budget, may imply a greater heat flow, as \citealt{li2018less} found in the case of Jupiter. This striking difference in heat flux suggests Uranus and Neptune have experienced different thermal history. One of the leading hypotheses for the low heat flux of Uranus is that convection in the deep interior is inhibited, preventing heat from being released as rapidly as it would in a fully adiabatic planet \citep{podolak1991models}. In addition, it has been proposed that the discrepancy in heat flow could be due to boundary layers in Uranus’ interior \citep{nettelmann2016uranus} or inhibited convection either in the deep interior \citep{podolak2019effect} or atmosphere and upper envelope \citep{markham2021constraining}. A low initial temperature for Uranus \citep{podolak1991uranus}, has been suggested as a potential reason why these planets evolved differently. Alternative explanations include giant impactor(s) prompting convective mixing in Neptune's interior or altering the atmospheric composition of both ice giants \citep{reinhardt2020bifurcation, morbidelli2012explaining, kegerreis2018consequences}.

 Indeed, the thermal evolution of these planets may depend on a variety of effects related to volatile enhancement in the envelopes and atmospheres. For example, as existing gravity data suggest different water mol fractions in the respective envelopes of these planets ($>$$10\%$ for Neptune and $<$$1\%$ for Uranus), \citet{bailey2021thermodynamically} suggested gradual demixing in the interior of Neptune but not Uranus may account for the discrepancy in heat flow. Another mechanism suggested to account for the ice giant heat flow discrepancy is release of latent heat in the atmosphere and upper envelope \citep{kurosaki2017acceleration}. In this model, pollution of the atmosphere and upper envelope by up to a $50\%$ mol fraction of heavy elements elevates planetary luminosity through latent heat release by condensation of molecules such as water, methane, and ammonia. The simultaneous effects of convective inhibition and latent heat were also considered by \citet{markham2021constraining} for their potential impact on planetary luminosity due to water and methane condensation assuming mol fractions of $5\%$ and $12\%$, respectively; while methane condensation can shorten the cooling timescale, water condensation will increase the required time to cool, both by up to $15\%$.  
 
In this work, we explore whether upsweep of icy material by the ice giants during their long-range Nice model migration was sufficient to produce the differences in volatile pollution necessary to account for these proposed effects. In order for these processes to explain the disparate heat of the ice giants, Uranus and Neptune must obtain different heavy element enhancements. We simulate the Nice Model migration and consider this late dynamical stage as a reason for the two planets' differences: Uranus and Neptune take different paths through the disk, accreting differing amounts of planetesimals of various compositions. While previous works \citep{tsiganis2005origin, batygin2010early, levison2011late} highlighted orbital sculpting of the outer solar system sculpting in an upheaval event, we focus here on the resultant impacts on the ice giants. We quantify differences in quantity and composition of accreted planetesimals, and the resulting potential for this dramatic phase of orbital upheaval to affect the composition and thermal evolution of Uranus and Neptune.

In Section \ref{sec:methods}, we discuss selecting initial conditions along with the construction of $N$-body Nice Model simulations. We normalize and categorize the varying accretion histories provided by our simulations in Section \ref{sec:results}. We comment on the quantity of heavy elements accreted to the envelope and atmospheres along with the potential for impacting the thermal evolution of Uranus and Neptune in Section \ref{sec:discussion}.

\section{Methods}
\label{sec:methods}

\begin{table*}[]
\caption{Initial conditions of the giant planets}
\label{tab:initialconditions}
\scriptsize
\begin{tabular}{lrrrrrrrr}
\hline
Body &Semimajor axis (au) &Eccentricity &Inclination ($\degree$) &True anomaly &$\varpi$ &$\Omega$ &Mass (M$_{\odot}$) &Radius (km)  \\
Jupiter &5.45003 &0.001 &0.001 &0 &$\pi$ &0 &9.54e-4 &6.99e4  \\
Saturn &7.17449 &0.001 &0.001 &$\pi$ &0 &0 &2.86e-4 &5.82e4  \\
Uranus &10.50767 &0.001 &0.001 &7$\pi$ /4 &$\pi$ &0 &4.50e-5 &2.50e4  \\
Neptune &17.6797 &0.001 &0.001 &5$\pi$ /16 &0 &0 &4.50e-5 &2.50e4 \\
\hline
\end{tabular}
\end{table*}

Simulations in this paper used the REBOUND $N$-body code \citep{rebound}. The simulations were integrated using IAS15, a 15th order Gauss-Radau integrator \citep{reboundias15}. We comment on this integrator choice in Appendix \ref{sect:int}.  We identify a set of initial conditions with Jupiter and Saturn in a 2:3 mean-motion resonance (MMR) in Section \ref{sec:initials}. A planetesimal disk was constructed to exclude dynamically cleared particles in Section \ref{sec:disk}. We discuss migration and damping forces in Sections \ref{sec:migration} and \ref{sec:damping} in lieu of including massive test particles to prompt a chaotic upheaval. A search for outcomes resembling our solar system is highlighted in Section \ref{sec:procedure}. The implementation of REBOUND's IAS15 integrator to preserve reproducibility is discussed in Section \ref{sec:modifications}. 

\subsection{Initial Conditions}\label{sec:initials}
Jupiter and Saturn were initially placed in a 2:3 MMR as in  \citet{masset2001reversing, morbidelli2007dynamics, pierens2008constraints}. Jupiter was initiated at 5.45 au and Saturn at 7.17 au following \citet{morbidelli2008late,tsiganis2005origin, batygin2010early}. The semimajor axes of the remaining two giant planets were randomly varied to find stable simulations. This was done by selecting from a normal distribution between 10 and 14.5 au for Uranus and 14.5 and 18 au for Neptune. We assumed as in \citet{morbidelli2007dynamics} that the planets came to these positions when the gas disk was present. While gas disk dispersal can itself cause dynamical instability, we are interested in Nice-model-like histories in which the planetary instability is delayed in time. We therefore integrate our simulations for 10$^{8}$ years and select initial configurations that are stable over this timescale.
 
\begin{figure*}[]
\includegraphics[height=6cm]{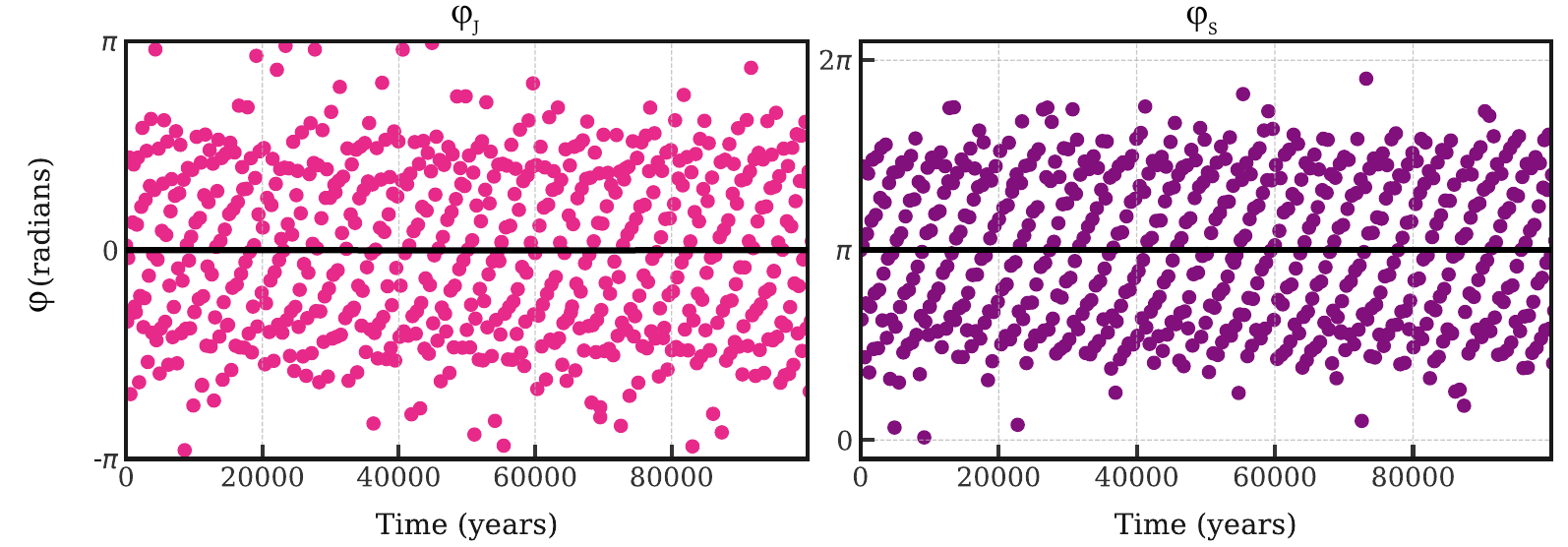} 
\caption{Jupiter and Saturn were initialized in 3:2 mean motion resonance, with libration angles $\varphi_{J} = 3\lambda_{S} - 2\lambda_{J}-\varpi_{J}$ and $\varphi_{S} = 3\lambda_{S} - 2\lambda_{J}-\varpi_{S}$ where $\lambda$ refers to the mean longitude and $\varpi$ is the longitude of pericenter. The angle $\varphi_{J}$ librates around 0 while $\varphi_{S}$ librates around $\pi$. Only the first $1\times10^{5}$ years are shown for clarity. The ability of the resonance to survive 10$^{8}$ years was used to determine the set of initial conditions used in this simulation (Table \ref{tab:initialconditions}), which we employ for all subsequent simulations in this work.  The relatively weak and noisy resonant behavior evident here results from inclusion of all four giant planets as well as the low initial eccentricities of Jupiter and Saturn.}
\label{fig:philibration}
\end{figure*}

 Because upheaval simulations are chaotic, a single set of stable initial conditions can be perturbed to generate a range of representative outcomes.  From the set described above, we pick the simulation in which the 3:2 mean-motion resonance angles for Jupiter and Saturn maintained the tightest resonant libration.  The four giant planets were given eccentricity and inclination of 0.001 as in \citet{tsiganis2005origin}. As illustrated in Figure \ref{fig:philibration}, even in the tightest resonant configuration found by this method, the resonance angles exhibit large libration amplitudes, which we interpret as coming from the planets' initially low eccentrities as well as perturbations from the other two giant planets. The final planetary initial conditions are shown in Table \ref{tab:initialconditions}.  We use these initial conditions for the giant planets in all subsequent simulations.  In what follows, we refer to the simulation containing only the giant planets initialized with these initial conditions and employing no additional forces as the ``base simulation."

\subsection{Planetesimal Disk}\label{sec:disk}
\begin{figure}[]
\begin{center}
\includegraphics[trim={1cm .45cm 1cm .45cm},clip,width=.6\textwidth]{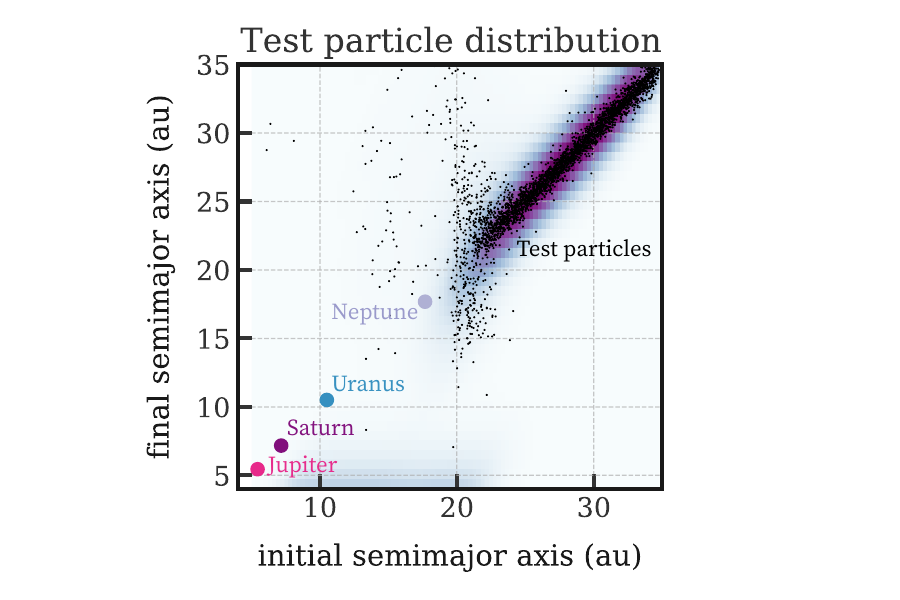}
\caption{Test particles were integrated with initial conditions (Section \ref{sec:disk}) to determine the placement of the planetesimal disc. Final vs initial semimajor axes of all test particles are shown after integration for $3$ million years. Planet initial positions are shown for this non-chaotic simulation. Test particles (black dots) were almost entirely removed inwards of $20$ au and thus this was chosen for the inner edge of the disk. Higher density is shown as darker purple.
}
\label{fig:particledisc}
\end{center}
\end{figure}

To determine the inner edge of the planetesimal disk of test particles, a suite of simulations were run with the base simulation, with the planetary initial conditions given in Table \ref{tab:initialconditions}. Test particles were added on circular orbits with $i = 0\degree$ and semi-major axes randomly drawn from a uniform distribution between 5 au and 35 au.  Particles were cleared by the stable orbits of the giant planets after integrating for 3 million years. The disk was almost completely cleared interior to 20 au and slightly perturbed interior to 23 au, but relatively unaffected beyond this point, as shown in Figure \ref{fig:particledisc}. Based on this result, the inner edge of the disk was placed at 20 au in all test cases. The outer edge of the disk was placed at 40 au to maximize potential for accretion by covering the full range that Neptune may roam into. Truncation of the solar system's planetesimal disk at $\sim$30 au has been suggested to explain the final semi-major axis of Neptune at the end of its outward migration \citep{tsiganis2005origin, levison2008origin}. The initial semi-major axes of accreted particles found in this work are presented in Section \ref{sec:results}, allowing for straightforward reanalysis of our results given any choice of truncation radius. 

Motivated by observations of protoplanetary disks \citep[e.g.][]{andrews2009protoplanetary}, we used a surface density distribution $\propto r^{-1}$ 
to replicate the expected density of solids assuming a Nice-model type evolution, consistent with \citet{morbidelli2007dynamics,thommes2008resonant, morbidelli2008late, batygin2010early}. In Section \ref{sec:origin}, we provide the percent of planetesimals accreted per 2 au band such that a reanalysis of our results can be performed with any surface density. 
To construct the modeled planetesimal disk, a rejection method was employed to choose a random semimajor axis for each test particle between $20$ and $40$ au. Since the area of the disk $\propto r^2$, our surface density distribution requires a probability $\propto r$ of choosing each randomly drawn semimajor axis. 

Planetesimals were initially given zero eccentricity and inclination.  We drew true anomaly, longitude of ascending node, and argument of pericenter randomly from a uniform distribution between 0 and 2$\pi$.
\subsection{Migration Force}\label{sec:migration}
As discussed in \citet{fernandez1984some}, exchange of orbital momentum between planets and planetesimals would cause Jupiter to migrate inwards and Saturn outwards. Since we model planetesimals as test particles for computational efficiency, we must impose this migration by hand. 
A velocity-dependent migration acceleration was added to the orbits of Jupiter and Saturn to move these planets to their current positions, following \citet{thommes2008resonant, morbidelli2005chaotic, levison2008origin}. As a result, Jupiter and Saturn leave their resonance, causing a system-wide upheaval. Our migration acceleration was derived from
\begin{equation}
    a(t) = a_{f}- \Delta a e^{-t/\tau},
    \label{Eq:migform}
\end{equation}
where $a$ is the semimajor axis and $\Delta a$ is defined as $a_{f} - a_{i}$, the final minus initial semimajor axes. The migration timescale is given by $\tau$ and $t$ is time in years. This was motivated to force the gas giants to a predetermined final semimajor axis, with migration decaying over time \citep[e.g.,][]{malhotra1993origin}. To implement this migration, we add an additional force on Jupiter and Saturn given by 
\begin{equation}
\dot{\vec{\mathbf{v}}} = \left(\frac{\vec{\mathbf{v}}}{2a}\right)\left(\frac{\Delta a_{J,S}}{\tau}\right)e^{-t/\tau}
\label{Eq:migrationforce}
\end{equation}
where the subscripts $J$ and $S$ apply to Jupiter and Saturn, $a$ and $\vec{\mathbf{v}}$ are the semimajor axis and velocity values at each timestep, and $\Delta a$ represents the desired total change in semimajor axis.

The form of Equation (\ref{Eq:migrationforce}) may be understood by considering a circular orbit.  
The change in semimajor axis required by Equation (\ref{Eq:migform}) is $\dot{a} = \frac{\Delta a}{\tau} e^{-\frac{t}{\tau}}$. For a circular orbit, Kepler's third law may be written 
$v = (GM_*/a)^{1/2}$ so that $\dot{v} = -(1/2)(GM_*/a^3)^{1/2} \dot a = -(v/2a)\dot a$,
where $M_{*}$ is the mass of the sun, $v$ is the velocity, and $G$ is the gravitational constant. Combining these expressions yields the planar components of Equation (\ref{Eq:migrationforce}).
The total change in semimajor axis was set such that Jupiter and Saturn would cease migration at their present day values. Migration out of the resonance triggers gravitational upheaval of the ice giants.  As the focus of our study was on the accretionary histories of Uranus and Neptune post-upheaval, a short migration timescale for Jupiter and Saturn was chosen for convenience, such that the orbital evolution of the four giant planets approximately matched those accepted within the Nice model \citep[e.g.][]{gomes2005origin}. Values for the constants are shown in Table \ref{tab:constants}. 
\vspace*{\fill}  

\subsection{Damping Force}\label{sec:damping}
Upheaval models rely on dynamical friction between the planets and planetesimals \citep{stewart1988evolution} to reduce the eccentricities of the ice giants from their upheaval values to those observed today.
Such damping is observed in previous works using a massive planetesimal disk (e.g. \citealt{tsiganis2005origin}, \citealt{batygin2010early}). 
Because computations using massive planetesimals are expensive, \cite{levison2008origin} used damping forces to replicate the effects of dynamical friction in their outer solar system upheaval model, which hosts only test particles. We take a similar approach here. After the initial chaotic period lasting $10^5$ years (Table \ref{tab:constants}), a damping force was added to the simulations, again such that we obtained a Nice-model type orbital evolution of the giant planets.

No force can change solely a planet's eccentricity, $e$, without also altering its semi-major axis, so we choose a damping force that primarily, though not exclusively, damps eccentricity.  This force (per mass) takes the form\footnote{This damping reproduces \citet{lee2002dynamics}, Figure 4 and is described in the REBOUND example ``Planetary migration in the GJ876 system (C)" (\url{https://rebound.readthedocs.io/en/latest/c_examples/planetary_migration/}).  Their added acceleration $\dot{\vec{\mathbf{v}}} = (2/3) \tau_e^{-1}(1-e^2)^{-1}\left[-\vec{\mathbf{v}} + a^{-1} \vec{\mathbf{h}}\times \hat{\mathbf{r}}\right]$ is equivalent to Equation (\ref{Eq:damping}) except for the factor of 2/3, which may be folded into $\tau_e$ without changing the functional form.}

\begin{equation}
    \dot{\vec{\mathbf{v}}} = -\frac{1}{\tau_e(1-e^2)}\left[\vec{\mathbf{v}} - \frac{h}{a}\hat{\mathbf{\theta}}\right]
    \label{Eq:damping}
\end{equation}
where $h$ is the magnitude of the angular momentum vector $\vec{\mathbf{h}} = \vec{\mathbf{r}} \times \vec{\mathbf{v}}$, and $\vec{\mathbf{r}}$ and $\vec{\mathbf{v}}$ are the position and velocity vectors measured with respect to the system's center of mass, with magnitudes $r$ and $v$ respectively.  The constant $\tau_e$ represents the eccentricity damping timescale and the semi-major axis $a$ is calculated in the center of mass frame. We refer to the unit vectors in the radial and azimuthal directions as $\hat{\mathbf{r}}$ and $\hat{\mathbf{\theta}}$.  We add the acceleration given in Equation \ref{Eq:damping} to our integration component-by-component in Cartesian coordinates, using the conversions $a = -\mu/(v^2 - 2\mu/r)$ and $e = |\vec{\mathbf{e}}|$, where
$\mu$ refers to $GM_{\rm tot}$, $M_{\rm tot}$ is the sum of the mass of the Sun and the mass of the planet undergoing damping, and 
the eccentricity vector $\vec{\mathbf{e}} = -\vec{\mathbf{h}}\times\vec{\mathbf{v}}/\mu - \hat{\mathbf{r}}$.
The origin of the acceleration in Equation (\ref{Eq:damping}) may be understood as follows.  For a Keplerian orbit, $\vec{\mathbf{v}} = (h/a)(1-e^2)^{-1}\left[e\sin{f} \hat{\mathbf{r}} + (1 + e\cos{f})\hat{\mathbf{\theta}}\right]$, where $f$ is the object's true anomaly  \citep{murray1999solar}.  Differentiating this expression with respect to time yields 
\begin{equation}
\vec{\mathbf{v}} = -\frac{\mu}{r^2}\hat{\mathbf{r}} + \frac{1}{2}\frac{\dot{a}}{a}\vec{\mathbf{v}} + \frac{\dot{e}}{e}\frac{1}{(1-e^2)}\left[\vec{\mathbf{v}}-\frac{h}{a}\hat{\mathbf{\theta}}\right] \;\;.  
\label{Eq:vdot}
\end{equation}
The first term on the right hand side of Equation (\ref{Eq:vdot}) is the gravitational acceleration due to the central mass.  If we take $\dot{a}/a = 0$ and $\dot{e}/e = \tau_e^{-1}$, we recover the added force in Equation (\ref{Eq:damping}).  These last choices are not self-consistent---the change in energy per mass with time, $\dot{C}$, resulting from the acceleration in Equation (\ref{Eq:damping}) is not zero and hence adding this force produces a non-zero $\dot{a}$.  However, it may be seen by computing $\dot{C} = \vec{\mathbf{v}} \cdot \dot{\vec{\mathbf{v}}}$ that the timescale $|C/\dot{C}| = |a/\dot{a}|$ on which this force changes the energy $C = -\mu/(2a)$ is longer than $|e/\dot{e}| \sim \tau_e$ by a factor of order $e^{-1}$. Qualitatively, this arises because $\vec{\mathbf{v}}-(h/a)\hat{\mathbf{\theta}}$ is approximately the orbit's epicyclic velocity.

\begin{wraptable}{l}{4cm}
\caption{Force constants for migration and damping}\label{tab:constants}
\scriptsize
\begin{tabular}{lrr}
\hline
$\Delta$ $a_{J}$ (au) &0.25 \\
$\Delta$ $a_{S}$ (au) &-2.36551 \\
$\tau_{migration}$ (years) &$\sim$1000 \\
$\tau_{e}$(years) &7.5e6 \\
Damping delay (years) &1e5 \\
\hline
\end{tabular}
\end{wraptable}

We comment that while \citet{fan2017simulations} find that integrating a full self-gravitating disk of planetesimals does not produce substantially different Nice-model-type outcomes, planetesimal-driven migration and dynamical friction does produce qualitative differences from the results we achieve using this damping term.  In particular, both pure planetesimal-driven migration \citep{fernandez1984some} and the late-stage planetesimal-driven stage of a Nice-model-like upheaval \citep{tsiganis2005origin} exhibit outward migration for the ice giants and Saturn, while Jupiter migrates inward. This behavior results from global exchange of planetesimals; this is not captured in eccentricity damping forces, which cause all planets to move inward.  Given the relatively short distance of this migration in the majority of our models, we do not expect this difference to substantially affect our results, but we return to this point in Section \ref{sec:caveats}. Dynamical friction and migration generated by planetesimals is also more stochastic than modeled here \citep{murray2006brownian,nesvorny2016neptune, hermosilloruiz2023}, but this behavior is unlikely to affect the rate of planetesimal collisions with the ice giants.  In short, the eccentricity damping force employed here serves the purpose of causing the planets to evolve qualitatively consistently with the Nice model. 

\begin{wraptable}{L}{3cm}
\begin{center}
\caption{Values for migration delays}\label{tab:migrationdelays}
\scriptsize
\begin{tabular}{lrr}
\hline
Simulation & Delay (years)\\
No Swap 1 &443 \\
No Swap 2 &392 \\
Swap 1 &265\\
Swap 2 &170 \\
\hline
\end{tabular}
\end{center}
\end{wraptable}
\subsection{Simulation procedure}\label{sec:procedure}
To reduce computational expense, we initially simulated the four giant planets without planetesimals with the above fictitious forces as in \citet{morbidelli2005chaotic, levison2008origin, thommes2008resonant} in order to determine appropriate migration and damping timescales. The time that the migration force was initiated was randomly selected between 100 and 1.5$\times10^{7}$ years, acting on the gas giants at different positions in their orbits. This served to generate a different chaotic simulation for each selected time. Selected times are shown in Table \ref{tab:migrationdelays}. As each of these simulations employed the same initial base simulation that was stable for $10^{8}$ years we can essentially access all phase space covered by the planets in just one orbital period. Thus simulations with shorter times before migration were preferred as there would be a more complete orbital evolution with less computational time. 

We developed a permissive criterion to determine if simulations outcomes were similar to the current solar system using the percent difference of the semimajor axes of each planet, defined as $a_{s}/a_{r}-1$ where $a_{s}$ is the final simulated semimajor axis and $a_{r}$ is the real present day axis. An acceptable simulation was defined as having an average percent difference between the four planets of less than $8\%$, with no individual planet to exceed a percent difference of $20\%$. Additionally, the eccentricity of each planet must be $<$$0.1$. 

\begin{figure*}[]
\includegraphics[width=\textwidth]{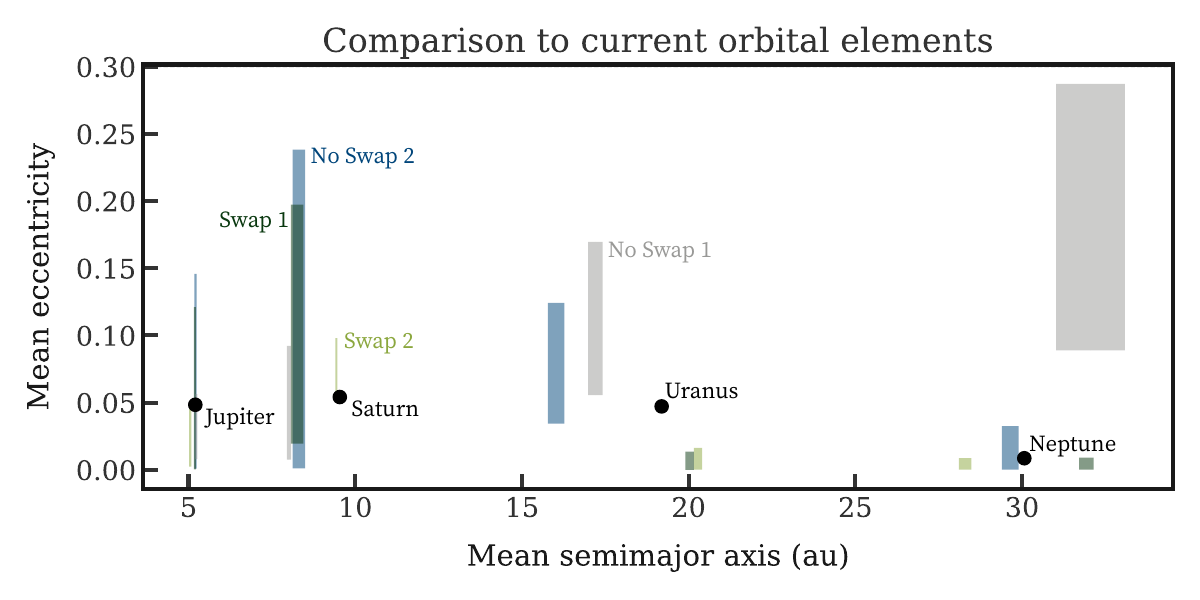}
\caption{Simulated mean semimajor axes and eccentricities for each planet (colored rectangles) compared with the corresponding present day values retrieved from JPL Horizons (black dots). Width and height of the rectangles correspond to the maximum and minimum values of the last $5$ million years of the four chosen runs. In two of the runs, Uranus started and ended interior to Neptune (No Swap 1 and 2) while in the other two Uranus started exterior to Neptune (Swap 1 and 2).
}
\label{fig:currentcompare}
\end{figure*}

With our selected timescales, $\sim$$2\%$ of runs had outcomes which aligned satisfactorily with the current positions of the ice giants according to our criteria as shown for four runs in Fig. \ref{fig:currentcompare}. The remaining $98\%$ had outcomes that did not remotely resemble that of the solar system: missing an ice giant or having one or more planets at $40+$ au were the most common alternatives, though there were simulations in which the gas giants were significantly disrupted from their orbits.  The ice giants swapped positions in approximately half of the accepted runs, in agreement with \citet{tsiganis2005origin}. For convenience, only four simulations were selected to integrate with test particles, two for each scenario. Full orbital evolutions for the four chosen runs can be seen in Fig. \ref{fig:allruns}. These results are consistent with \citet{tsiganis2005origin} and \citet{batygin2010early}. 

\subsection{Reproducibility Modifications}\label{sec:modifications}
In $N$-body integrations with many massless test particles, it is more computationally efficient to run numerous simulations in parallel and compile the data from all massless particles at the end. Thus, our integrator needed to be reproducible: for any given set of initial conditions, the giant planets must follow the same orbital evolution regardless of the number of massless bodies present. For non-chaotic simulations, this is generally not an issue, as the massless bodies by default do not affect massive bodies. However, due to numerical precision, the addition of these bodies can change the roundoff error of the force calculations. In chaotic systems, these errors propagate to the extent that the outcome of the simulation is substantially affected. 

In order to maintain reproducibility in the face of roundoff errors, the following integration scheme was implemented. The planetary initial condition archive file was loaded into a first simulation object to create a planet-only simulation, which was then copied. Randomly generated massless test particles were added to the copy, now referred to as the particle simulation. The planet-only simulation is run for one timestep and then copied, creating a third simulation object. The particle simulation is then run to that same time, then the test particles are copied over into the copied simulation. The copied simulation now becomes the new particle simulation. The old particle simulation is freed, and the process repeats. This ensures that the planetary motions are only dependent on bodies with mass and are not being affected by roundoff errors in the timesteps due to calculations of interactions.
\section{Results}
\label{sec:results}
We integrated each of the four chosen simulations with test particles and determine the number of collisions in Section \ref{sec:collisions}. We normalized the simulated test particles to a standard disk mass in Section \ref{sec:norm} to obtain the mass increase for each ice giant, given in Section \ref{sec:massincrease}. We determined the formation location of the accreted planetesimals in Section \ref{sec:origin}. In the following analysis, Uranus (the final interior ice giant) is always associated with blue, and Neptune (the final exterior ice giant) with purple.
\subsection{Collisions}\label{sec:collisions}
\begin{figure*}[]
\begin{center}
\includegraphics[width=\textwidth]{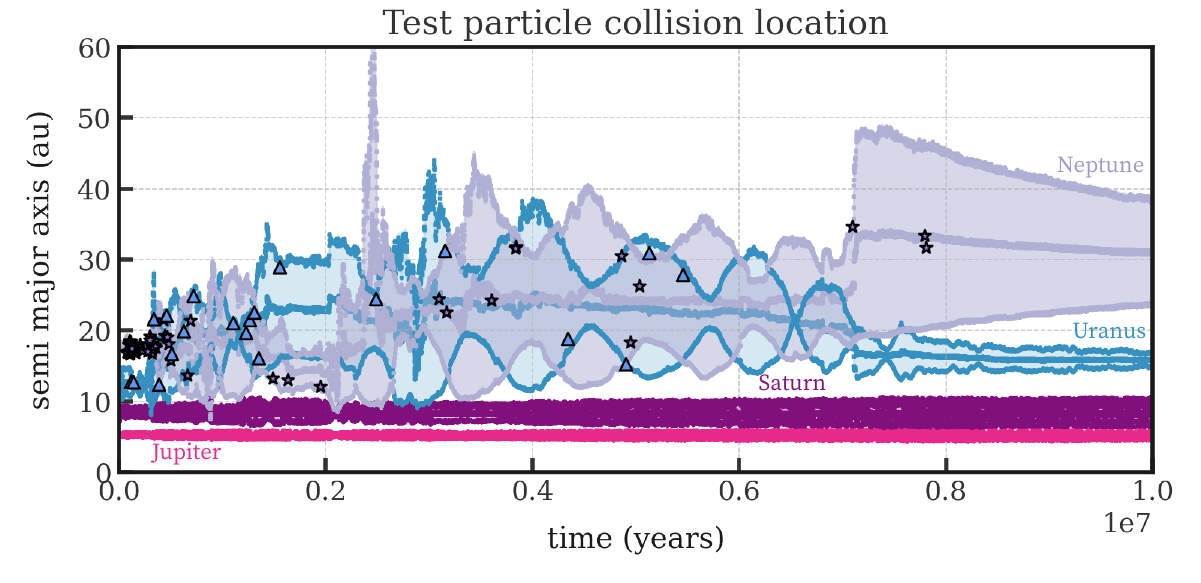}
\caption{Semimajor axis ($a$), apocenter distance $a(1+e)$, and pericenter distance $a(1-e)$ versus time of Jupiter (pink), Saturn (dark purple), Uranus (blue), and Neptune (light purple) along with the time and location of collisions with test particles for Uranus (blue triangles) and Neptune (purple stars) during the chaotic period. In this scenario (No Swap 1), the ice giants did not end with swapped positions: Neptune began and ended exterior to Uranus. Neptune was responsible for $\sim$$80\%$ of the total test particle collisions. }
\label{fig:particlecollisions}
\end{center}
\end{figure*}

Each of the four chosen simulations was integrated with $\sim$100,000 test particles in order to get sufficient statistics on planetesimal accretion. One outcome is shown in Figure \ref{fig:particlecollisions}; analogous plots for all four chosen simulations and plots showing the evolution of the planetary orbits over longer timescales are available in Appendix \ref{App:figs}. 
From this, the number of test particle collisions with each planet was acquired for each scenario. Collisions were detected with REBOUND's Line detection module and resolved with a custom collision resolve function. This recorded the identifying hash of the colliding test particle and planet before removing the test particle from the simulation.
Collisions were predominantly due to particles initially located at the inner edge of the disk, with the outermost ice giant accreting the majority, as shown in Figure \ref{fig:particlelocations}. For all simulations, the majority of collisions occured at the beginning of the simulation, with collision frequency decreasing with time.
\subsection{Normalization}\label{sec:norm}
 To convert from the number of collisions that occur in our simulations to the physical collision rate, we normalize 
 assuming an initial surface density in planetesimals of 0.25 g/cm$^{2}$ at 30 au, i.e. surface density $\Sigma_p = \Sigma_{p,0}(r/30 {\rm au})^{-1}$ with $\Sigma_{p,0}=$ 0.25 g/cm$^{2}$. As a result we are simulating a disk in which the total mass of planetesimals is $\sim$$35 
 M_{\oplus}$. This is comparable to the minimum mass solar nebula of \citet{hayashi1981structure}, using the density for rock+ice over the same disk radii, and the disk mass assumed in \citet{desch2007mass}. \citet{morbidelli2008late} report that more massive disks did not provide qualitatively similar outcomes to our solar system. We note that the chaotic outcomes of this simulations imply that a large number of simulations would be required to explore the full phase space of outcomes for a given disk mass. We used a disk surface density with a shallower power law than those employed in \citet{hayashi1981structure} and \citet{desch2007mass}. For a planetesimal surface density $\propto r^{-1}$, our computed rates of mass accretion onto the planets are linearly proportional to surface density at 30 au used for normalization.  Steeper profiles would preferentially weight collisions with planetesimals originating near the inner edge of the disk (cf. Figure \ref{fig:particlelocations}).

 \subsection{Mass increase due to accretion}\label{sec:massincrease}
Table \ref{tab:massinc} summarizes the percent mass increase due to accretion for three scenarios which each consider different regions where accreted planetesimals may be considered fully ablated and well mixed (discussed in depth in Section \ref{sec:discussion}). The first scenario assumes the accretion is well mixed throughout the entire planet (taken as $10^{29}$ g) while the second assumes planetesimals are only mixed throughout the envelope. The third assumes (small) planetesimals ablate in the upper envelope and are prevented from mixing with the remainder of the envelope due to inhibited convection.  Entries are presented as the total final mass in the considered region after accretion divided by the initial mass. Considered as a percentage of the total planet mass, the greatest increase was $0.35\%$ for Neptune in a non-swapping run, with Uranus only accreting an additional $0.1\%$ of its mass. 

Uranus averaged a $0.1\%$ mass increase in all four scenarios. For scenarios in which the ice giants swapped, Neptune averaged a $0.04\%$ mass increase, while for the non-swapping scenarios Neptune's $0.35\%$ and $0.1\%$ increases were larger. A larger suite of simulations would be required to fully explore the level of dispersion in outcomes. Table \ref{tab:massinc} also provides the enhancement factor by mass of the planets' gaseous envelopes, which we take to comprise 10\% of the planets by mass. 

\begin{table*}[]
\centering
\caption{Mass gained due to planetesimal accretion with respect to three regions: the full planet, the envelope, and the upper envelope to the 100 bar level (total final mass/initial mass) }
\label{tab:massinc}
\scriptsize
\begin{tabular}{lrrrrrrrr}
\hline
Simulation &\multicolumn{2}{c}{Swap 1} &\multicolumn{2}{c}{Swap 2} &\multicolumn{2}{c}{No Swap 1} &\multicolumn{2}{c}{No Swap 2} \\
Planet &Uranus &Neptune &Uranus &Neptune &Uranus &Neptune &Uranus &Neptune \\
Total planet mass &1.0011 &1.0004 &1.0010 &1.0003 &1.0010 &1.0035 &1.0009 &1.0010 \\
Envelope mass &1.011 &1.004 &1.010 &1.003 &1.010 &1.035 &1.009 &1.010 \\
100 bar level mass &7800 &2500 &7300&2400 &7300 &27000 &7000 &7300 \\
\hline
\end{tabular}
\end{table*}

\vspace*{\fill}   
\subsection{Origin of Accreted Planetesimals}\label{sec:origin}

\begin{figure}[]
\centering
\includegraphics[trim={.1cm 0 .1cm 0},clip,width=.48\textwidth]{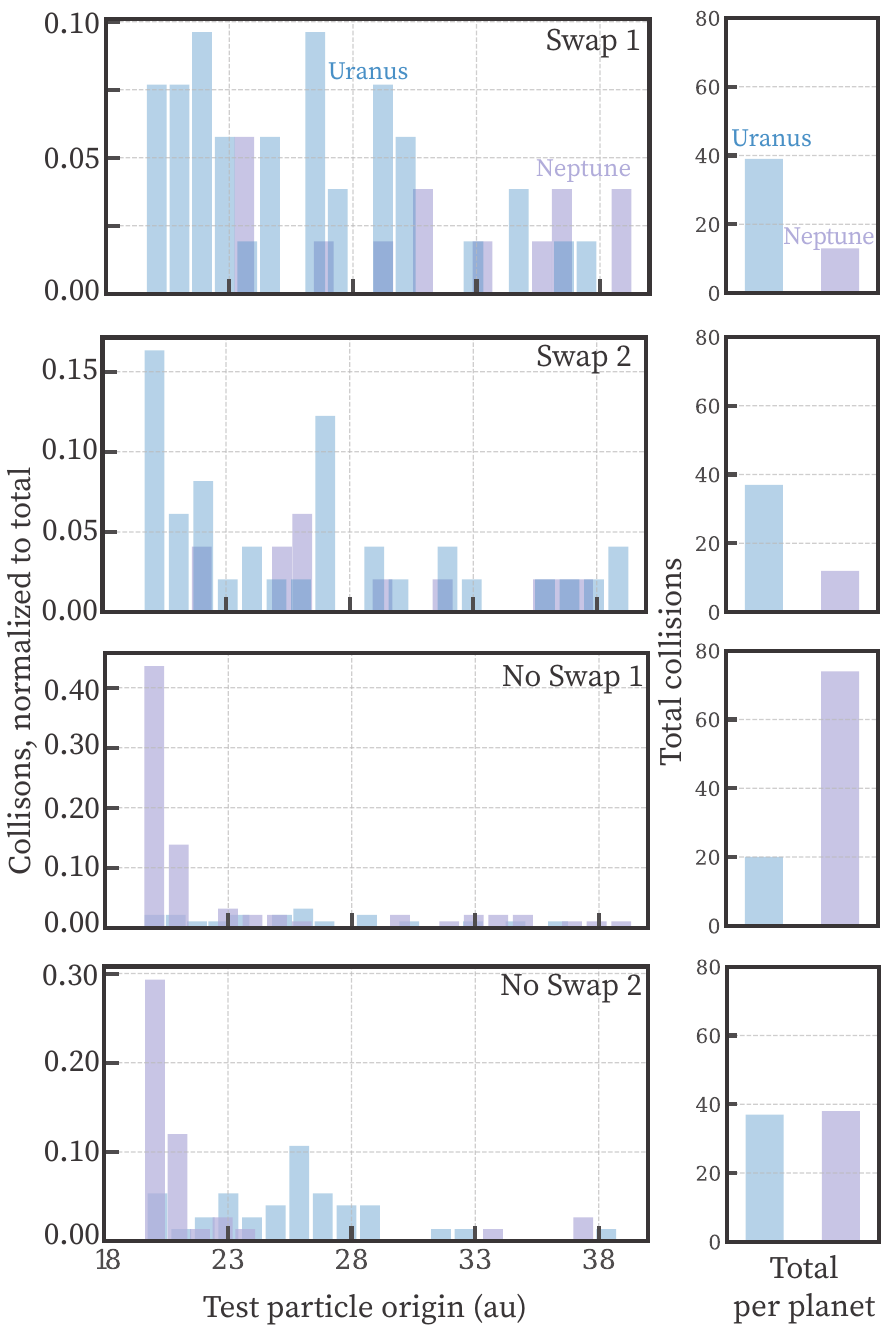} 
\caption{The left column shows initial locations of accreted test particles in the disk and the planet they collided with. The right column gives the total collisions for Uranus (blue) and Neptune (purple), with the planets identified by their locations in their final configuration. The first two rows show simulations in which Uranus initially is exterior to Neptune (Swaps 1\&2) whereas the bottom two rows show scenarios in which Neptune starts and ends exterior to Uranus (No Swaps 1\&2). Collisions are normalized to the total number undergone by both ice giants in each simulation. The majority of  particles were accreted by the initially outermost ice giant from the inner edge of the disk. The remainder of the accreted particles originated from all regions of the disk with little variation.}
\label{fig:particlelocations}
\end{figure} 

\begin{figure*}[]
\centering
\includegraphics[width=\textwidth]{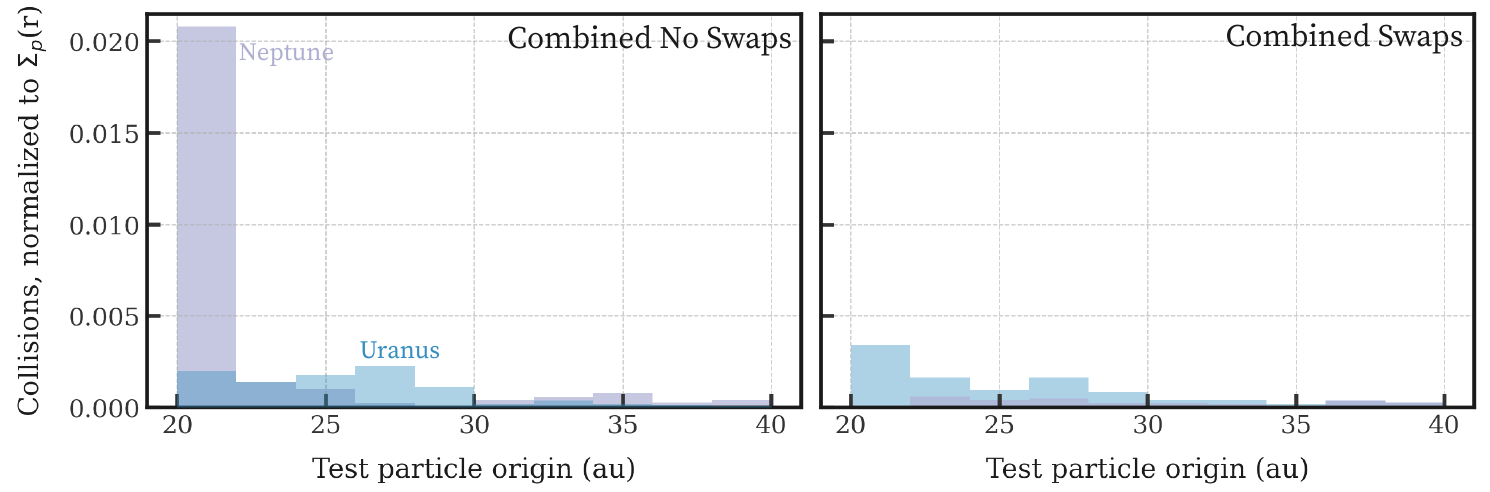}
\caption{Fraction of particles initially located in each 2 au radial bin that ultimately collide with Neptune (purple) and Uranus (blue). This provides the probability of collision for a given initial test particle semimajor axis, for any initial surface density. The left plot shows the combined statistics for No Swaps 1\&2 while the right plot shows the same for Swaps 1\&2. Uranus and Neptune  are identified by their locations in their final configuration.}
\label{fig:surfacedensityparticlelocations}
\end{figure*} 
To investigate the origins of planetesimals that ultimately collide with the ice giants, we display the initial semi-major axes for these planetesimals in Figure \ref{fig:particlelocations}. In swapping simulations, Uranus received the most collisions from planetesimals initialized between $\sim$$20-24$ au. Neptune varied between Swaps 1 and 2, accreting mostly planetesimals originating in the region $\sim$$36-40$ au in Swap 1 and $\sim$$25-29$ in Swap 2. In both, the inner 2-3 au of the planetesimal disk was only accreted by Uranus. In the non-swapping simulations, Uranus predominately accreted planetesimals initialized between 24 and 28 au while Neptune mainly collided with planetesimals originating between 20 and 24 au. Notably, in non-swapping simulations, Uranus accretes a significant portion of planetesimals formed between 25-38 au while beginning and ending interior to Neptune. Throughout all simulations, the average initial semimajor axis of test particles which collided with Uranus was $\sim$$26$ au. For Neptune, this value was $30$ au for swapping scenarios and $23$ au for non-swaps: Neptune accreted test particles formed farther out in the disk when it was shielded by an initially exterior Uranus. The initially exterior ice giant accreted the majority of the inner test particles, leaving only those farther out for the initially interior planet. In swapping simulations, the inner $2-3$ au of the disk was reserved exclusively for Uranus. However, while Neptune dominated accretion in this region in non-swapping scenarios, Uranus was able to accrete in this area. Thus in both swapping and non swapping variants, Uranus accretes planetesimals from $20-23$ au while Neptune only does so when it is initially exterior. A discussion of the icelines present in this disk can be found in Section \ref{sec:nitrogen}.

Both swapping simulations accreted planetesimals from a similar distribution of initial radii, as did both non-swapping simulations, as seen in Figure \ref{fig:particlelocations}. We used this to determine the impact of our assumed surface density on the radial collision distribution by summing the collisions of the two swapping, and, separately, the two non-swapping simulations. This was normalized to the initial planetesimal surface density and is displayed in Figure \ref{fig:surfacedensityparticlelocations}. This provides a way to translate the number of collisions found to any initial planetesimal surface density. The accretionary trends apparent in Figure \ref{fig:particlelocations} remain visible in Figure \ref{fig:surfacedensityparticlelocations}: the initial chosen surface density does not greatly influence this result. \pagebreak

\section{Discussion}\label{sec:discussion}
During the ice giants' Nice Model migration, we find that these planets experienced a late accretion of icy planetesimals equivalent to an envelope mass fraction of up to $\sim$$3.5\%$. This phase of solar system formation would have constituted an extreme bombardment event for Uranus and Neptune: assuming the majority of the mass is in km-sized bodies, one of these planets would have collided with up to $3$ planetesimals per hour during the first million years of the giant planets' orbital instability. Further, the initially exterior ice giant accretes more planetesimals. We now address the possible relevance of this substantial impactor flux to the early thermal evolution of Uranus and Neptune. 

As discussed in Section \ref{sec:introduction}, Uranus has a much smaller present-day observed heat flow than Neptune, despite the fact that these planets share many outward similarities. We compare our simulations with published models to evaluate whether the substantial volatile accretion evident in our simulations could have caused Uranus to experience rapid cooling compared to Neptune at early stages of solar system formation after the dispersal of the gas disk, providing an explanation for the difference in heat flow. The depth at which planetesimal ablation occurs determines which part of the planet is affected, so we consider the full envelope separately from the upper envelope and atmosphere.
\subsection{Comparison to previous work}
The extent of accretion undergone by the ice giants during a Nice model upheaval was considered in \citet{matter2009calculation}. As mentioned in Section \ref{sec:introduction}, we employed a collision detection module within REBOUND to directly simulate collisions between planets and planetesimals instead of determining collisions retroactively using collision probabilities as utilized by \citet{matter2009calculation}. In all cases, we find less mass accreted than this previous work by a factor of 2 to 10 times. It should be noted that while the total planetesimal disk mass used by \citet{matter2009calculation} is comparable to that used in our work, the disk used in this previous study begins at 15.5 au while our disk begins at 20 au, as we excluded these dynamically cleared particles assumed to have incorporated into the planet during earlier planet formation as discussed in Section \ref{sec:disk}. As the majority of accretion in our simulations was due to planetesimals at the inner edge of our disk, we can speculate that including planetesimals interior to 20 au would substantially increase total accretion relative to the case studied here. Given this difference, we cannot rule out the possibility that the N-body and analytical models give different results. However, we do reproduce the finding that the initially exterior ice giant undergoes the most accretion.

\subsection{Heavy element pollution in Uranus and Neptune's envelopes}\label{sec:envelope}
To satisfy gravity observations, models require metal mole fractions of $\geq$ 
 $10\%$ and $<$$1\%$ in the envelopes of Neptune and Uranus, respectively \citep{bailey2021thermodynamically}. Contamination of the envelopes of these planets by icy planetesimals depends on planetesimal size as this controls where ablation occurs. \citet{pinhas2016efficiency} find that icy impactors with radii $\sim$$1$ km fall through the atmosphere and ablate almost entirely by the $1000$-bar level inside the planet due to drag and thermal effects. Given this finding, it is plausible that $100$-km-sized planetesimals would also ablate in the envelope, which extends to the $10^{5}$-bar level \citep{nettelmann2013new}.  

Motivated by these ablation levels for large planetesimals, we first consider a scenario in which the accreted material is mixed by convection throughout a hydrogen-dominated envelope with $10\%$ of the planet mass (Table \ref{tab:initialconditions}) \citep{hubbard1980structure}. 

 As mentioned in Section \ref{sec:introduction}, one or more giant impacts have previously been proposed to account for the observed differences in Uranus and Neptune. We can consider the accretion examined in our work to be caused by any plausible planetesimal size-mass distribution; namely, we can consider the full mass increase and corresponding change in envelope compositions as being delivered either over time by many small impactors or all at once by a $\sim$$ 1/20 M_{\oplus}$ impactor, the maximum total accreted mass.
 We employ the full mass increase for each planet over the full simulated time, as documented in Table \ref{tab:massinc}, to obtain the mass gained in volatiles (also referred to as ``metals" or ``water"). While it is unrealistic to assume the planetesimals are fully water, it provides a limiting case and allows for direct comparison to previously computed models which utilized this same assumption, such as \citet{kurosaki2017acceleration}. It is straightforward to translate the mass accreted that we find to any planetesimal composition (see Section \ref{sec:nitrogen} for an example). 
 
 We calculated water mole fractions in the envelopes due to accretion, with the assumption that the remaining mass of the envelope is a H$_{2}$-He mixture in solar proportions (mean molar mass $2.3$ g/mol). 

Because Uranus and Neptune swapped position in some simulations but not others, this begs the question of how these two different migratory outcomes relate to the amount of material accreted. In all four simulations, Uranus obtained a $\sim$$0.1\%$ water mole fraction in its envelope. As seen in Figure \ref{fig:particlelocations}, Neptune's accretion varied substantially.
In the simulation with highest accretion (No Swap 1), Neptune obtained a volatile mass fraction of $\sim$$3\%$ in its envelope, corresponding to a $\sim$$0.5\%$ water mole fraction. In swapping simulations, Neptune only obtained a water mole fraction of $\sim$$0.04\%$.

Our enhancements for Uranus were found to be below the limiting value of $\sim$$1\%$, which, long-term settling notwithstanding, agree with current gravity observations that suggest Uranus is more centrally condensed \citep[e.g.][]{helled2010interior, nettelmann2013new, podolak1995comparative}. However, even the case of greatest accretion simulated for Neptune is insufficient to provide the heavy element enrichment of Neptune observed today. Accordingly, the substantial difference in heavy element enrichment of the envelope must have been obtained by Uranus and Neptune through a different process than disparate accretion of a ``late veneer" as explored in this work. While volatile enrichment of the envelope during the Nice model migration was not sufficient to account for the entire heavy element content of Neptune's envelope, this late-stage accretion was enough to increase the mean molecular weight of the envelope by up to $\sim$$3\%$.

\subsection{Heavy element pollution in the atmosphere and upper envelope}\label{sec:atmosphere}

As discussed in Section \ref{sec:introduction}, condensation of water and methane in the atmosphere and upper envelope may play a role in the thermal evolution of the ice giants. \citet{kurosaki2017acceleration} find that a $50\%$ water mol fraction in a young Uranus' atmosphere and upper envelope is sufficient to cool the planet to its observed flux by present day. Additionally, they find that this luminosity is almost three orders of magnitude higher than that due to a $45\%$ mol fraction. \citet{markham2021constraining} consider the effects on planetary luminosity of stable stratification alongside latent heat release; they show that while methane condensation increases luminosity, water condensation causes a longer cooling timescale.

Accordingly, we consider the possibility that the late accretion quantified in this work enriched the atmospheres and upper envelopes of the ice giants at early times, leading to transient but important changes in cooling. We find that in swapping simulations, Uranus is enhanced more than Neptune; we sought to quantify the effect of the excess atmospheric pollution of Uranus. Following the logic of \citet{kurosaki2017acceleration}, for the purposes of this discussion, we assume the mass of the impactors is entirely in water. Because the mean molecular weights of ammonia and methane are similar to water, the following conclusions can be loosely applied to ``ices" as a whole. We further follow their approach by considering a region of the planet's atmosphere and upper envelope which extends to the $100$ bar level. 

Small water-ice planetesimals will ablate at or above this level; in order to consider this scenario, we must examine the potential for the existence of these small planetesimals. While ALMA observations show $\sim$$1$ mm planetesimals (often referred to as ``pebbles") present in gas disks with a collective mass comparable to that of the minimum mass solar nebula \citep[e.g.][]{andrews2009protoplanetary, andrews2015observations}, it is unclear if particles of this size would still be present after disk dispersal. If they were to remain, collisional grinding would likely reduce the population of the largest-size pebbles on a timescale of $t = (4/3)(\rho_{\rm int} s/\Sigma) \Omega^{-1}$. 
We estimate that a surface density of pebbles $\Sigma \gtrsim 5\times 10^{-6}$ g cm$^{-2}$ at $25$ au would result in destruction by collisional grinding over timescales less than $1$ Myr, where we have used an internal pebble density   

$\rho_{\rm int} = 2$ g cm$^{-3}$, and $\Omega$ is the angular orbital velocity of a particle at 25 au. Alternatively, if small planetesimals did not survive disk dispersal, they could be formed through a collisional cascade generated by a reservoir of $1$ km bodies. In a classic collisional cascade, the number of particles at a given size, $N(s)$, has size distribution $dN/ds \propto s^{-q}$, so that the mass at a given size is proportional to $s^{4-q}$.  We take $q$=3.5. For the surface density discussed in Section \ref{sec:norm}, the time between collisions for $1$ km sized objects is $\sim$$20$ million years, meaning that as long as the Nice Model upheaval occurs at least $20$ Myr after disk dispersal a collisional cascade is likely established, well within the $\sim$$800$ Myr delay proposed by \citet{gomes2005origin}. If this scenario were to occur,  we can expect a small planetesimal surface density of $2.5\times10^{-4}$ g cm$^{-2}$ in mm-sized objects. 

For the remainder of this section, we consider a fiducial surface density of $2.5\times10^{-4}$ g cm$^{-2}$ of 1 mm planetesimals to get a basic understanding of the potential implications.  This size is small enough to ablate entirely in the atmosphere and upper envelope by the $1$ $\mu$bar level \citep{moses1992meteoroid}. An additional ablation model would be required to determine the maximum planetesimal size to be ablated by the $100$ bar level, which could be a topic of future work. 

We normalized simulated collisions following Section \ref{sec:norm} and consider the mass increase due to accretion in comparison to the size of the atmosphere. We focus on the first million years of accretion, when impact rates were highest. Collision rates peaked at $10^{15}$ per hour (Neptune, No Swap 1), while Uranus averaged $10^{14}$ per hour during this time across all four simulations. We can expect these small impactors to settle to the envelope on timescales of approximately $100$ years \citep{mordasini2014grain}. Moreover, while not well constrained, the convective overturn timescale of Neptune has been estimated at $100$ years \citep{hubbard1984planetary}. Thus regardless of whether the impactors were fully or only mostly ablated, they should remain in the atmosphere and upper envelope at minimum on the order of $100$ years.

 We estimated the mass of the atmosphere and upper envelope ($\sim$$10^{22}$g) to be a small fraction of the planet mass, calculated as $4\pi r_{p}^{2} \gamma H \rho_{p}$ where $r_{p}$ and  $\rho_{p}$ are the planetary radius and density, respectively. We determined $\rho_{p}$ at the base of the upper envelope as $P/c_{s}^{2}$, with $P$ as the pressure of 100 bar. The sound speed $c_{s}$ is calculated as $\sqrt{RT/M}$ where $R$ is the gas constant, $T$ is the temperature ($350$ K \citep{mousis2021situ}), and $M$ the mean molar mass of $2.3$ g/ml (a solar H$_{2}$-He mix). We assumed an adiabatic region with $\gamma \approx 7/5$, the adiabatic index of H$_{2}$, as $100$ bar is below the generally expected radiative-convective boundary and therefore assumed to be convective. The scale height $H$ was found as $c_{s}^{2}r_{p}^2/(G M_{p})$, where $M_{p}$ is the planet mass. Using the mass of the atmosphere and upper envelope, we determined the water mole fraction of this region due to planetesimal accretion.  

The convective mixing timescale is long enough in swapping simulations for Uranus' atmosphere and upper envelope to accrete $\sim$$6\times10^{-5}$ its original mass, obtaining a water mol fraction of $\sim$$ 8\times10^{-6}$. Neptune gains up to $0.002$ of the mass of its atmosphere and upper envelope in this time, corresponding to a water mol fraction of $\sim$$10^{-4}$. 

However, it is not certain that this region would be mixed on this $100$ year timescale: convective upwelling or inhibited convection may keep heavy elements aloft. Interior models of the ice giants can include layers of inhibited convection beginning at the $200$-bar level \citep{leconte2017condensation}. For this case, we consider total accretion over the first million years. Even though the metallicities that we compute are large enough that they will likely be subject to overturn instabilities, this atmospheric enrichment is directly considered, following \citet{kurosaki2017acceleration}. Enhancement is much greater than when considering a $100$-year mixing timescale, with Neptune acquiring up to $25 \times$ its atmospheric and upper envelope mass. (Note that this enhancement is much smaller than the enhancements provided in Table \ref{tab:massinc}, which include all of the accreted planetesimal mass, because we have limited ourselves to the minority of the mass contained in small pebbles.) For both ice giants in all modeled scenarios, the volatile mole fractions in the atmosphere and upper envelope were $>$$0.2$, with a maximum value of $0.8$ for Neptune and $0.5$ for Uranus. This implies a heavily volatile-enriched atmosphere and upper envelope. 

This enhancement is sufficient to alter the thermal evolution of the ice giants according to published models; substantially so in the case of \citep{kurosaki2017acceleration}. As mentioned in Section \ref{sec:introduction}, \citet{markham2021constraining} determine that condensation of methane and water can alter the cooling timescale by up to $15\%$, either accelerating or lengthening the required time. We find much higher mol fractions than their assumed $5\%$ for methane and $12\%$ for water; these were their highest studied mol fractions and they found that the change in cooling timescale increased with heavy element concentration. We can speculate that this trend will continue. Without constraining the compositions of water and methane in the disk at this time, it is unclear whether we would expect our simulated planets to undergo an increased (mostly water accreted) or reduced (mostly methane accreted) cooling timescale. However, we found that the atmosphere and upper envelope of the ice giants were flooded with volatiles at early times, dominating the composition of this region.

 For Uranus to experience an accelerated cooling timescale early in its evolution, we require a swapping simulation in which pebbles are dominantly composed of methane, increasing the planet's luminosity at earlier times, thus accelerating its cooling. We found that Uranus incurred an atmospheric water mol fraction $\sim$$2$ times that of Neptune's in swapping simulations. Conversely, in non-swapping simulations, Neptune obtained $1.5$ times the atmospheric water mole fraction of Uranus. In a non-swapping scenario, we would require mainly water to be accreted, causing Neptune's cooling timescale to increase compared to that of Uranus. In either case, accretion is significant enough that cooling timescales of both planets would likely simultaneously be affected. If the relative amounts of water and methane in pebbles could be determined during this time period, we could theorize on the starting configuration of the ice giants. 
 We comment that while pebble accretion will increase grain input in the upper atmosphere, \citet{mordasini2014grain} find that in the outer radiative zone this increase in dust does not significantly increase the grain opacity; further, the gas opacity more strongly effects the thermal evolution \citep{lunine1989effect}.
 
 We note that the details of the atmosphere are a key aspect in dictating how a planet cools over time \citep[e.g.,][]{fortney2011self}. \citet{fortney2011self} computed atmospheric models of Uranus and Neptune in order to determine the thermal evolution, and while they were able to match Neptune's known luminosity, they were unable to do so for Uranus. This model focused on the atmospheric boundary conditions, and highlights the need for new calculations of atmospheric boundary conditions for ice giant evolution models, including the possibility of planetesimal dissolution in the atmosphere. While a higher opacity may slow cooling, the details of atmospheric opacities are complex. We may expect that the influx of material associated with planetesimal dissolution would increase the planet's opacity, increasing the solar energy input while decreasing the energy output, but more updated work is needed to better understand this feedback.

\subsection{$N_{2}$ and Kr icelines}\label{sec:nitrogen}
We now consider the effect of icelines within the planetesimal disk on the atmospheric enhancement of Uranus and Neptune due to accreted planetesimals. We note that as Uranus and Neptune began formation in different locations of the disk before undergoing the upheaval described in this work, they may have had different atmospheric compositions initially. This model only calculates the enrichment undergone during late-stage accretion and does not account for any initial metallicity difference. As discussed in Section \ref{sec:atmosphere}, in the case of a 100-year convective mixing timescale, the metallicity increase due to this accretion is small to moderate compared to any initial enhancement, and the initial atmospheric composition would be required to predict what we observe today. However, in the case of inhibited convection preventing mixing throughout the envelope, enhancement would be dominated by the accretion explored in this work, and the initial atmospheric composition can be neglected. This is true even if the metallicity of Neptune's atmosphere was previously enhanced to today's observed value. 

As highlighted in Figure \ref{fig:particlecollisions}, the radial source of the accreted planetesimals varied between swapping and non-swapping simulations. We considered whether this difference in planetesimal source populations provides a measureable compositional signature of whether the ice giants swapped orientation during a dynamical upheaval. Such utility would require a compositional tracer or tracers that vary across the radial region of the disk where the planetesimals originate. The N$_{2}$ iceline at $\sim$$26$ au and the krypton (Kr) iceline at $\sim$$22$ au \citep{2019AJ....158..194O} offer perhaps the most promising possibilities. This model is based on a water snowline of $\sim$$2$ au and a corresponding midplane temperature  of 140K, with temperature $\propto$ $r^{-0.65}$. Interior to these ice lines, N$_{2}$ and Kr are primarily in gaseous form, while exterior to the ice lines, they are primarily components of solid planetesimals and hence are available to be accreted by the ice giants during the collisions described in this work. In particular, in non-swapping scenarios we may expect krypton (and to a lesser extent, nitrogen) accretion in Uranus but little to none in Neptune, as Uranus accretes planetesimals predominately from exterior to $22$ au while Neptune accretes from interior to this radius. In swapping scenarios we may expect enhancement of krypton in both planets while nitrogen accretion is restricted to Uranus, as both planets accrete planetesimals from the $22-26$ au region but Neptune's accretion is minimal exterior to this. As no carbon icelines exist within the extent of our disk, carbon is accreted in all simulations and thus provides a baseline for comparison.

The post-formation enrichments discussed below are linked to the initial orbital rank of the planets and thus can be used as a constraint in this respect.
To test this idea, we assigned compositions to the planetesimals in our simulations. Planetesimals were assumed to have compositions equivalent to the solar nebula with abundances following \citet{2019AJ....158..194O}, with mass split evenly \citep{greenberg1998making} between silicates and volatiles such as CO, CO$_{2}$, ethane, nitrogen, ammonia, water, and trace noble gases. For the subset of scenarios discussed in Section \ref{sec:atmosphere}, where volatile enrichment dominates the atmosphere and upper envelope, we find our prediction is true. As the amount of accretion varies per simulation, we define $f_{i}$ as the ratio of $i$:C accreted by the planet to the solar value of $i$:C, where $i$ refers to Kr or N. For krypton, Uranus obtains an approximately solar enhancement in all simulations ($f_{Kr} \sim$$0.9-1.1$). In swapping scenarios, Neptune has a solar enrichment ($f_{Kr}\sim$$1.1$) while in non-swapping cases this enhancement is subsolar ($f_{Kr}\sim$$0.2-0.3$).

When considering nitrogen enrichment, we find subsolar ratios in all cases for Neptune ($f_{N}\sim$$0.1-0.3$). In swapping scenarios, Uranus obtains a $\sim$$2$ times solar enhancement while in non-swapping cases enrichment is subsolar: $f_{N}\sim$$0.6-0.8$. It is likely that convective mixing throughout the envelopes of the ice giants has occurred since this epoch and these minor relative enrichments will be unable to be detected in their current atmospheres; however, the resulting enrichment may be more apparent on the satellites of the ice giants. 

\subsection{Caveats to the model}\label{sec:caveats}
Our models employed several simplifications. As discussed in Section \ref{sec:introduction}, the orbital evolution of the giant planets during the Nice model evolution has been heavily studied; we sought to replicate this while reducing computational expense by avoiding the use of massive test particles. Similar to existing Nice model simulations, we assumed fictional forces to provoke dynamical instability and to simulate damping caused by dynamical friction. As discussed in Sections \ref{sec:migration} and \ref{sec:damping}, our migration force causes Jupiter and Saturn to divergently migrate, in agreement with models of planetesimal-driven migration \citep[e.g.,][]{fernandez1984some}. However, we neglect dynamical friction by massive test particles in favor of our eccentricity damping force, which reduces the ice giants semimajor axes along with their eccentricities. While reduction of semimajor axis would not occur with real dynamical friction, our eccentricity force was made as weak as possible to reduce this effect while still providing needed damping. The majority of collisions occured during the first million years of the simulation when semimajor axes, and thus the eccentricity damping force, were lower. Further, the semimajor axis space covered by the giant planets during their simulated migration is sufficiently limited such that they are consistent with published examples of the Nice model migration. Accordingly, a fictional eccentricity force is sufficient in place of massive test particles for the purposes of this study.

Moreover, only four different orbital evolution scenarios were simulated in depth. Even with massless test particles, the computational expense of simulations with tens of thousands of test particles is high, and we sought to investigate a few outcomes in detail rather than obtain the complete range of possibilities. We see large  variations in outcomes within our small sample size: in No Swap 1, total accretion for Neptune was a factor of 4 higher than either planet in any other scenario. These simulations provide a starting point to understand accretionary differences between the two scenarios. We would require a large suite each of numerous orbital evolution scenarios to obtain greater statistics. Future works could explore the full range of possibilities in greater detail. 

We did not consider the scenario where there are initially three ice giants \citep[e.g.,][]{batygin2011instability, nesvorny2011young}. It may be fair to speculate that--as in our two-planet simulations--the initially outermost ice giant would accrete the majority of the planetesimals. If this were the ice giant to be ejected, this could leave less planetesimals for Uranus and Neptune, preventing as extreme an impact to their thermal evolution and atmospheric compositions. Conversely, if the initially innermost ice giant were ejected, accretion could be similar to the two ice giant case. However, the three-planet scenario could benefit from future work, particularly as observable properties may be concerned.

In addition, we only considered one planetesimal surface density. Proposed planetesimal distributions are varied \citep[e.g.][]{WEIDENSCHILLING2011671, schlichting2013initial, johansen2014protostars}, thus we provide estimates at both ends of the size spectrum. Our results can be scaled to fit any surface density $\propto r^{-1}$. Further, a different functional form of the radial dependence can be investigated using our results normalized to the initial disk surface density, available in Figure \ref{fig:surfacedensityparticlelocations}. Ablation depth is dependent on impactor size; we require an additional ablation calculation to determine the size of impactor that would ablate by the $100$-bar level.
\section{Conclusion}

We have estimated the amount of planetesimal accretion by Uranus and Neptune during the Nice model migration, investigating how this late stage of accretion may have impacted the present-day properties of these planets according to existing models. We carried out a suite of direct $N$-body orbital simulations and found that the ice giants undergo an extreme bombardment period lasting a million years with mass accretion rates corresponding to collision rates of up to $3$ planetesimals with $1$-km radius per hour. This estimate takes into account a standard assumption for the surface density of $\propto r^{-1}$ with 0.25 g/cm$^{2}$ at 30 au.  

We have found that Uranus and Neptune accrete differing amounts of volatiles during the Nice model migration, and specific outcomes vary depending on whether the planets switch orbital rank. The initially exterior ice giant accretes the most planetesimals. In simulations where Uranus is initially exterior to Neptune, both ice giants accrete planetesimals formed between 22-26 au. In simulations where Uranus begins and ends interior to Neptune, Neptune accretes the majority of its planetesimals from the inner disk. 
 
 When considering the possible role of inhibited convection or convective upwelling in maintaining an elevated fraction of heavy elements in the outer region of the planet, the atmosphere and upper envelope of the ice giants down to the $\sim$$100$ bar level can be compositionally dominated by the volatiles accreted during this period. Transient water mol fractions up to $80\%$ are estimated from our orbital simulations. This estimate does not account for the possibility of hydrodynamic mixing instabilities. If this material exits the atmosphere and upper envelope on an estimated $100$-year convective timescale, then the metallicity of the planet's outer region is not sufficiently enhanced to alter rates of early thermal evolution. We note that even if inhibited convection or convective upwelling prevent mixing in the upper envelope and atmosphere during the early thermal evolution of the planets, mixing over the age of the solar system may have erased enhancements from this bombardment event that could be observed today.

Future work constraining convective mixing timescales and regions of inhibited convection in the envelopes of the ice giants is needed to understand the effect of volatile accretion on the thermal evolution.  Based on the results of these simulations, we suggest water mol fractions $>$$0.5$ in the upper envelope may have been temporarily sustained if convection was inhibited below the atmosphere, as suggested by \citet{leconte2017condensation}. Crucially, even a brief atmospheric enrichment has the potential to affect the long-term planetary evolution, as \citet{kurosaki2017acceleration} find a sharp increase of multiple orders of magnitude in planetary luminosity when comparing between a $45\%$ and $50\%$ volatile mol fraction in the atmosphere. Accordingly, this work motivates further investigation of the effect of latent heat release and inhibited convection at water mol fractions above $50\%$. The thermodynamic histories and interior structures of Uranus and Neptune have depended on a variety of complicated factors over the lifetime of the solar system. We have established here that these planets experienced substantial accretion of volatiles from the massive primordial disk during their long-range outward migration. Accordingly, this late stage of accretion by Uranus and Neptune deserves further consideration as a potentially important influence on the observable properties of these planets.

\begin{acknowledgements}
This work was made possible with funding from the Heising-Simons Foundation 51 Pegasi b Postdoctoral Fellowship in Planetary Astronomy and the Undergraduate Research in Science and Technology award from UC Santa Cruz. We acknowledge use of the lux supercomputer at UC Santa Cruz, funded by NSF MRI grant AST 1828315. We thank Jonathan Fortney and an anonymous reviewer for helpful comments.  
\end{acknowledgements}

\bibliography{reboundbib}{}

\begin{thebibliography}{}
\expandafter\ifx\csname natexlab\endcsname\relax\def\natexlab#1{#1}\fi
\providecommand{\url}[1]{\href{#1}{#1}}
\providecommand{\dodoi}[1]{doi:~\href{http://doi.org/#1}{\nolinkurl{#1}}}
\providecommand{\doeprint}[1]{\href{http://ascl.net/#1}{\nolinkurl{http://ascl.net/#1}}}
\providecommand{\doarXiv}[1]{\href{https://arxiv.org/abs/#1}{\nolinkurl{https://arxiv.org/abs/#1}}}

\bibitem[{Andrews(2015)}]{andrews2015observations}
Andrews, S.~M. 2015, Publications of the Astronomical Society of the Pacific, 127, 961

\bibitem[{Andrews {et~al.}(2009)Andrews, Wilner, Hughes, Qi, \& Dullemond}]{andrews2009protoplanetary}
Andrews, S.~M., Wilner, D., Hughes, A., Qi, C., \& Dullemond, C. 2009, The Astrophysical Journal, 700, 1502

\bibitem[{Bailey \& Stevenson(2021)}]{bailey2021thermodynamically}
Bailey, E., \& Stevenson, D.~J. 2021, The Planetary Science Journal, 2, 64

\bibitem[{Baines \& Smith(1990)}]{baines1990atmospheric}
Baines, K.~H., \& Smith, W.~H. 1990, Icarus, 85, 65

\bibitem[{Batygin \& Brown(2010)}]{batygin2010early}
Batygin, K., \& Brown, M.~E. 2010, The Astrophysical Journal, 716, 1323

\bibitem[{Batygin {et~al.}(2011)Batygin, Brown, \& Betts}]{batygin2011instability}
Batygin, K., Brown, M.~E., \& Betts, H. 2011, The Astrophysical Journal Letters, 744, L3

\bibitem[{Connerney {et~al.}(1991)Connerney, Acuna, \& Ness}]{connerney1991magnetic}
Connerney, J., Acuna, M.~H., \& Ness, N.~F. 1991, Journal of Geophysical Research: Space Physics, 96, 19023

\bibitem[{Connerney {et~al.}(1987)Connerney, Acuña, \& Ness}]{connerney1987magnetic}
Connerney, J. E.~P., Acuña, M.~H., \& Ness, N.~F. 1987, Journal of Geophysical Research: Space Physics, 92, 15329, \dodoi{https://doi.org/10.1029/JA092iA13p15329}

\bibitem[{Conrath {et~al.}(1987)Conrath, Gautier, Hanel, Lindal, \& Marten}]{Conrath1987helium}
Conrath, B., Gautier, D., Hanel, R., Lindal, G., \& Marten, A. 1987, Journal of Geophysical Research: Space Physics, 92, 15003

\bibitem[{Conrath {et~al.}(1991{\natexlab{a}})Conrath, Gautier, Lindal, Samuelson, \& Shaffer}]{Conrath1991helium}
Conrath, B., Gautier, D., Lindal, G., Samuelson, R., \& Shaffer, W. 1991{\natexlab{a}}, Journal of Geophysical Research: Space Physics, 96, 18907

\bibitem[{Conrath {et~al.}(1991{\natexlab{b}})Conrath, Pearl, Appleby, Lindal, Orton, \& Bezard}]{conrath1991thermal}
Conrath, B., Pearl, J., Appleby, J., {et~al.} 1991{\natexlab{b}}, Uranus, 204

\bibitem[{Conrath {et~al.}(1989)Conrath, Flasar, Hanel, Kunde, Maguire, Pearl, Pirraglia, Samuelson, Gierasch, Weir, {et~al.}}]{conrath1989infrared}
Conrath, B., Flasar, F., Hanel, R., {et~al.} 1989, Science, 246, 1454

\bibitem[{De~Wit \& Seager(2013)}]{de2013constraining}
De~Wit, J., \& Seager, S. 2013, Science, 342, 1473

\bibitem[{Desch(2007)}]{desch2007mass}
Desch, S. 2007, The Astrophysical Journal, 671, 878

\bibitem[{Fan \& Batygin(2017)}]{fan2017simulations}
Fan, S., \& Batygin, K. 2017, The Astrophysical Journal Letters, 851, L37

\bibitem[{Fernandez \& Ip(1984)}]{fernandez1984some}
Fernandez, J., \& Ip, W.-H. 1984, Icarus, 58, 109

\bibitem[{Fortney {et~al.}(2011)Fortney, Ikoma, Nettelmann, Guillot, \& Marley}]{fortney2011self}
Fortney, J.~J., Ikoma, M., Nettelmann, N., Guillot, T., \& Marley, M. 2011, The Astrophysical Journal, 729, 32

\bibitem[{Frelikh \& Murray-Clay(2017)}]{frelikh2017formation}
Frelikh, R., \& Murray-Clay, R.~A. 2017, The Astronomical Journal, 154, 98

\bibitem[{Gomes {et~al.}(2005)Gomes, Levison, Tsiganis, \& Morbidelli}]{gomes2005origin}
Gomes, R., Levison, H.~F., Tsiganis, K., \& Morbidelli, A. 2005, Nature, 435, 466

\bibitem[{Greenberg(1998)}]{greenberg1998making}
Greenberg, J.~M. 1998, Astronomy and Astrophysics, v. 330, p. 375-380 (1998), 330, 375

\bibitem[{Guillot \& Gautier(2015)}]{Guillot_2015}
Guillot, T., \& Gautier, D. 2015, Giant Planets (Elsevier), 529–557, \dodoi{10.1016/b978-0-444-53802-4.00176-7}

\bibitem[{Hayashi(1981)}]{hayashi1981structure}
Hayashi, C. 1981, Progress of Theoretical Physics Supplement, 70, 35

\bibitem[{Helled {et~al.}(2010)Helled, Anderson, Podolak, \& Schubert}]{helled2010interior}
Helled, R., Anderson, J.~D., Podolak, M., \& Schubert, G. 2010, The Astrophysical Journal, 726, 15

\bibitem[{Helled \& Bodenheimer(2014)}]{helled2014formation}
Helled, R., \& Bodenheimer, P. 2014, The Astrophysical Journal, 789, 69

\bibitem[{Helled \& Fortney(2020)}]{helled2020interiors}
Helled, R., \& Fortney, J.~J. 2020, Philosophical Transactions of the Royal Society A, 378, 20190474

\bibitem[{{Hermosillo Ruiz} {et~al.}(2023){Hermosillo Ruiz}, {Lau}, \& {Murray-Clay}}]{hermosilloruiz2023}
{Hermosillo Ruiz}, A., {Lau}, H., \& {Murray-Clay}, R. 2023, \mnras, submitted

\bibitem[{Hubbard {et~al.}(1995)Hubbard, Podolak, Stevenson, {et~al.}}]{hubbard1995interior}
Hubbard, W., Podolak, M., Stevenson, D., {et~al.} 1995, Neptune and Triton, 109

\bibitem[{Hubbard(1984)}]{hubbard1984planetary}
Hubbard, W.~B. 1984, New York

\bibitem[{Hubbard \& MacFarlane(1980)}]{hubbard1980structure}
Hubbard, W.~B., \& MacFarlane, J.~J. 1980, Journal of Geophysical Research: Solid Earth, 85, 225

\bibitem[{Jacobson(2014)}]{jacobson2014orbits}
Jacobson, R. 2014, The Astronomical Journal, 148, 76

\bibitem[{Johansen {et~al.}(2014)Johansen, Blum, Tanaka, Ormel, Bizzarro, Rickman, Beuther, Klessen, Dullemond, Henning, {et~al.}}]{johansen2014protostars}
Johansen, A., Blum, J., Tanaka, H., {et~al.} 2014

\bibitem[{Karkoschka \& Tomasko(2011)}]{karkoschka2011haze}
Karkoschka, E., \& Tomasko, M.~G. 2011, Icarus, 211, 780

\bibitem[{Kegerreis {et~al.}(2018)Kegerreis, Teodoro, Eke, Massey, Catling, Fryer, Korycansky, Warren, \& Zahnle}]{kegerreis2018consequences}
Kegerreis, J.~A., Teodoro, L., Eke, V., {et~al.} 2018, The Astrophysical Journal, 861, 52

\bibitem[{Kurosaki \& Ikoma(2017)}]{kurosaki2017acceleration}
Kurosaki, K., \& Ikoma, M. 2017, The Astronomical Journal, 153, 260

\bibitem[{Lambrechts \& Johansen(2012)}]{lambrechts2012rapid}
Lambrechts, M., \& Johansen, A. 2012, Astronomy \& Astrophysics, 544, A32

\bibitem[{Leconte {et~al.}(2017)Leconte, Selsis, Hersant, \& Guillot}]{leconte2017condensation}
Leconte, J., Selsis, F., Hersant, F., \& Guillot, T. 2017, Astronomy \& Astrophysics, 598, A98

\bibitem[{Lee \& Peale(2002)}]{lee2002dynamics}
Lee, M.~H., \& Peale, S.~J. 2002, The Astrophysical Journal, 567, 596

\bibitem[{Levison {et~al.}(2011)Levison, Morbidelli, Tsiganis, Nesvorn{\`y}, \& Gomes}]{levison2011late}
Levison, H.~F., Morbidelli, A., Tsiganis, K., Nesvorn{\`y}, D., \& Gomes, R. 2011, The Astronomical Journal, 142, 152

\bibitem[{Levison {et~al.}(2008)Levison, Morbidelli, VanLaerhoven, Gomes, \& Tsiganis}]{levison2008origin}
Levison, H.~F., Morbidelli, A., VanLaerhoven, C., Gomes, R., \& Tsiganis, K. 2008, Icarus, 196, 258

\bibitem[{Li {et~al.}(2018)Li, Jiang, West, Gierasch, Perez-Hoyos, Sanchez-Lavega, Fletcher, Fortney, Knowles, Porco, {et~al.}}]{li2018less}
Li, L., Jiang, X., West, R., {et~al.} 2018, Nature Communications, 9, 3709

\bibitem[{Lindal(1992)}]{lindal1992atmosphere}
Lindal, G.~F. 1992, Astronomical Journal (ISSN 0004-6256), vol. 103, March 1992, p. 967-982., 103, 967

\bibitem[{Loewenstein {et~al.}(1977{\natexlab{a}})Loewenstein, Harper, \& Moseley}]{loewenstein1977effective}
Loewenstein, R., Harper, D., \& Moseley, H. 1977{\natexlab{a}}, The Astrophysical Journal, 218, L145

\bibitem[{Loewenstein {et~al.}(1977{\natexlab{b}})Loewenstein, Harper, Moseley, Telesco, Thronson~Jr, Hildebrand, Whitcomb, Winston, \& Stiening}]{loewenstein1977far}
Loewenstein, R., Harper, D., Moseley, S., {et~al.} 1977{\natexlab{b}}, Icarus, 31, 315

\bibitem[{Lunine {et~al.}(1989)Lunine, Hubbard, Burrows, Wang, \& Garlow}]{lunine1989effect}
Lunine, J.~I., Hubbard, W., Burrows, A., Wang, Y.-P., \& Garlow, K. 1989, Astrophysical Journal, Part 1 (ISSN 0004-637X), vol. 338, March 1, 1989, p. 314-337. Research supported by the Alfred P. Sloan Foundation., 338, 314

\bibitem[{Malhotra(1993)}]{malhotra1993origin}
Malhotra, R. 1993, Nature, 365, 819

\bibitem[{Markham \& Stevenson(2021)}]{markham2021constraining}
Markham, S., \& Stevenson, D. 2021, The Planetary Science Journal, 2, 146

\bibitem[{Masset \& Snellgrove(2001)}]{masset2001reversing}
Masset, F., \& Snellgrove, M. 2001, Monthly Notices of the Royal Astronomical Society, 320, L55

\bibitem[{Matter {et~al.}(2009)Matter, Guillot, \& Morbidelli}]{matter2009calculation}
Matter, A., Guillot, T., \& Morbidelli, A. 2009, Planetary and Space Science, 57, 816

\bibitem[{Morbidelli \& Crida(2007)}]{morbidelli2007dynamics}
Morbidelli, A., \& Crida, A. 2007, icarus, 191, 158

\bibitem[{Morbidelli \& Levison(2008)}]{morbidelli2008late}
Morbidelli, A., \& Levison, H. 2008, Physica Scripta, 2008, 014028

\bibitem[{Morbidelli {et~al.}(2005)Morbidelli, Levison, Tsiganis, \& Gomes}]{morbidelli2005chaotic}
Morbidelli, A., Levison, H.~F., Tsiganis, K., \& Gomes, R. 2005, Nature, 435, 462

\bibitem[{Morbidelli {et~al.}(2012)Morbidelli, Tsiganis, Batygin, Crida, \& Gomes}]{morbidelli2012explaining}
Morbidelli, A., Tsiganis, K., Batygin, K., Crida, A., \& Gomes, R. 2012, Icarus, 219, 737

\bibitem[{Mordasini(2014)}]{mordasini2014grain}
Mordasini, C. 2014, Astronomy \& Astrophysics, 572, A118

\bibitem[{Moses(1992)}]{moses1992meteoroid}
Moses, J.~I. 1992, Icarus, 99, 368

\bibitem[{Mousis {et~al.}(2021)Mousis, Atkinson, Ambrosi, Atreya, Banfield, Barabash, Blanc, Cavali{\'e}, Coustenis, Deleuil, {et~al.}}]{mousis2021situ}
Mousis, O., Atkinson, D.~H., Ambrosi, R., {et~al.} 2021, Experimental astronomy, 1

\bibitem[{Murray \& Dermott(1999)}]{murray1999solar}
Murray, C.~D., \& Dermott, S.~F. 1999, Solar system dynamics (Cambridge university press)

\bibitem[{Murray-Clay \& Chiang(2006)}]{murray2006brownian}
Murray-Clay, R.~A., \& Chiang, E.~I. 2006, The Astrophysical Journal, 651, 1194

\bibitem[{Nesvorn{\`y}(2011)}]{nesvorny2011young}
Nesvorn{\`y}, D. 2011, The Astrophysical Journal Letters, 742, L22

\bibitem[{Nesvorn{\`y} \& Vokrouhlick{\`y}(2016)}]{nesvorny2016neptune}
Nesvorn{\`y}, D., \& Vokrouhlick{\`y}, D. 2016, The Astrophysical Journal, 825, 94

\bibitem[{Nettelmann {et~al.}(2013)Nettelmann, Helled, Fortney, \& Redmer}]{nettelmann2013new}
Nettelmann, N., Helled, R., Fortney, J., \& Redmer, R. 2013, Planetary and Space Science, 77, 143

\bibitem[{Nettelmann {et~al.}(2016)Nettelmann, Wang, Fortney, Hamel, Yellamilli, Bethkenhagen, \& Redmer}]{nettelmann2016uranus}
Nettelmann, N., Wang, K., Fortney, J.~J., {et~al.} 2016, Icarus, 275, 107

\bibitem[{{{\"O}berg} \& {Wordsworth}(2019)}]{2019AJ....158..194O}
{{\"O}berg}, K.~I., \& {Wordsworth}, R. 2019, \aj, 158, 194, \dodoi{10.3847/1538-3881/ab46a8}

\bibitem[{Pearl \& Conrath(1991)}]{pearl1991albedo}
Pearl, J., \& Conrath, B. 1991, Journal of Geophysical Research: Space Physics, 96, 18921

\bibitem[{Pierens \& Nelson(2008)}]{pierens2008constraints}
Pierens, A., \& Nelson, R.~P. 2008, Astronomy \& Astrophysics, 482, 333

\bibitem[{Pinhas {et~al.}(2016)Pinhas, Madhusudhan, \& Clarke}]{pinhas2016efficiency}
Pinhas, A., Madhusudhan, N., \& Clarke, C. 2016, Monthly Notices of the Royal Astronomical Society, 463, 4516

\bibitem[{Podolak {et~al.}(2019)Podolak, Helled, \& Schubert}]{podolak2019effect}
Podolak, M., Helled, R., \& Schubert, G. 2019, Monthly Notices of the Royal Astronomical Society, 487, 2653

\bibitem[{Podolak {et~al.}(1991{\natexlab{a}})Podolak, Hubbard, \& Stevenson}]{podolak1991uranus}
Podolak, M., Hubbard, W., \& Stevenson, D. 1991{\natexlab{a}}, Uranus,  Univ. of Arizona Press

\bibitem[{Podolak {et~al.}(1991{\natexlab{b}})Podolak, Hubbard, Stevenson, {et~al.}}]{podolak1991models}
Podolak, M., Hubbard, W., Stevenson, D., {et~al.} 1991{\natexlab{b}}, Uranus, 1, 29

\bibitem[{Podolak {et~al.}(1995)Podolak, Weizman, \& Marley}]{podolak1995comparative}
Podolak, M., Weizman, A., \& Marley, M. 1995, Planetary and Space Science, 43, 1517

\bibitem[{{Rein} \& {Liu}(2012)}]{rebound}
{Rein}, H., \& {Liu}, S.~F. 2012, \aap, 537, A128, \dodoi{10.1051/0004-6361/201118085}

\bibitem[{{Rein} \& {Spiegel}(2015)}]{reboundias15}
{Rein}, H., \& {Spiegel}, D.~S. 2015, \mnras, 446, 1424, \dodoi{10.1093/mnras/stu2164}

\bibitem[{{Rein} {et~al.}(2019){Rein}, {Hernandez}, {Tamayo}, {Brown}, {Eckels}, {Holmes}, {Lau}, {Leblanc}, \& {Silburt}}]{reboundmercurius}
{Rein}, H., {Hernandez}, D.~M., {Tamayo}, D., {et~al.} 2019, \mnras, 485, 5490, \dodoi{10.1093/mnras/stz769}

\bibitem[{Reinhardt {et~al.}(2020)Reinhardt, Chau, Stadel, \& Helled}]{reinhardt2020bifurcation}
Reinhardt, C., Chau, A., Stadel, J., \& Helled, R. 2020, Monthly Notices of the Royal Astronomical Society, 492, 5336

\bibitem[{Schlichting {et~al.}(2013)Schlichting, Fuentes, \& Trilling}]{schlichting2013initial}
Schlichting, H.~E., Fuentes, C.~I., \& Trilling, D.~E. 2013, The Astronomical Journal, 146, 36

\bibitem[{Soderlund {et~al.}(2013)Soderlund, Heimpel, King, \& Aurnou}]{soderlund2013turbulent}
Soderlund, K., Heimpel, M., King, E., \& Aurnou, J. 2013, Icarus, 224, 97

\bibitem[{Sromovsky {et~al.}(2011)Sromovsky, Fry, \& Kim}]{sromovsky2011methane}
Sromovsky, L., Fry, P., \& Kim, J.~H. 2011, Icarus, 215, 292

\bibitem[{Stanley \& Bloxham(2004)}]{stanley2004convective}
Stanley, S., \& Bloxham, J. 2004, Nature, 428, 151

\bibitem[{Stanley \& Bloxham(2006)}]{stanley2006numerical}
---. 2006, Icarus, 184, 556

\bibitem[{Stewart \& Wetherill(1988)}]{stewart1988evolution}
Stewart, G.~R., \& Wetherill, G.~W. 1988, Icarus, 74, 542

\bibitem[{Thommes {et~al.}(2008)Thommes, Bryden, Wu, \& Rasio}]{thommes2008resonant}
Thommes, E.~W., Bryden, G., Wu, Y., \& Rasio, F.~A. 2008, 398, 315

\bibitem[{Tsiganis {et~al.}(2005)Tsiganis, Gomes, Morbidelli, \& Levison}]{tsiganis2005origin}
Tsiganis, K., Gomes, R., Morbidelli, A., \& Levison, H.~F. 2005, Nature, 435, 459

\bibitem[{Tyler {et~al.}(1989)Tyler, Sweetnam, Anderson, Borutzki, Campbell, Eshleman, Gresh, Gurrola, Hinson, Kawashima, {et~al.}}]{tyler1989voyager}
Tyler, G., Sweetnam, D., Anderson, J., {et~al.} 1989, Science, 246, 1466

\bibitem[{Weidenschilling(2011)}]{WEIDENSCHILLING2011671}
Weidenschilling, S. 2011, Icarus, 214, 671, \dodoi{https://doi.org/10.1016/j.icarus.2011.05.024}

\bibitem[{Wetherill(1967)}]{wetherill1967collisions}
Wetherill, G. 1967, Journal of Geophysical Research, 72, 2429

\end{thebibliography}
\bibliographystyle{aasjournal}
\appendix
\section{Integrator Selection}\label{sect:int}
 We sought a capable integrator for a chaotic simulation which could accurately resolve close encounters. We implemented a selection scheme for the IAS15 \citep{reboundias15} and Mercurius integrators \citep{reboundmercurius} which involved varying the timestep and switchover parameters of Mercurius and comparing to the IAS15 outcome. We desired our outcomes to be independent of integrator; however, we found that in this chaotic scenario Mercurius and IAS15 yielded different results. This type of problem is known to be difficult for symplectic integrators and makes for an interesting test case.
 
\begin{figure}[b]
\begin{center}
\includegraphics[width=\textwidth]{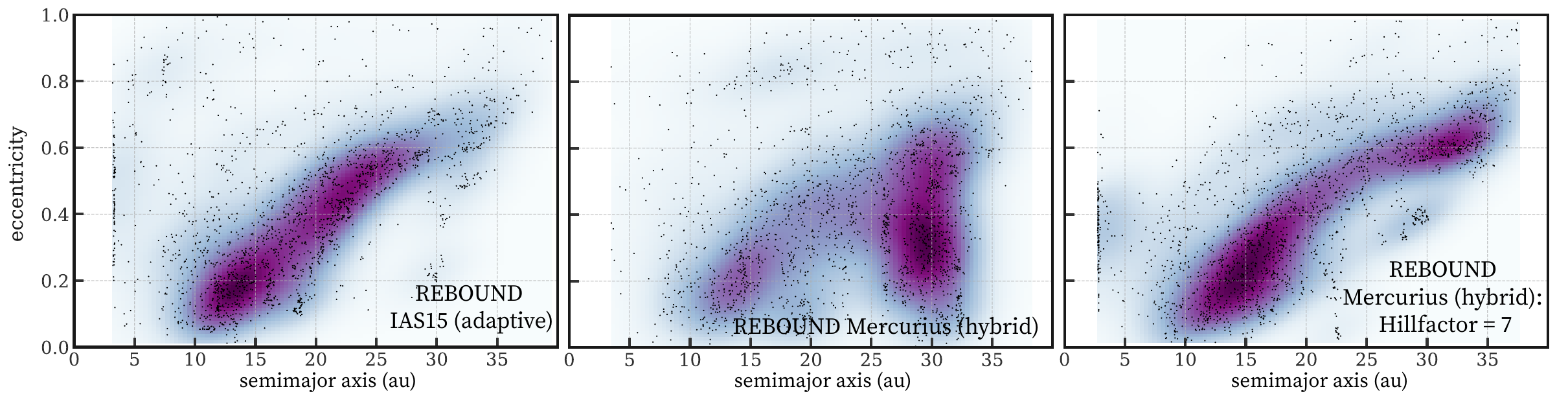} 
\end{center}
\caption{Semimajor axis-eccentricity density plots for Neptune representing the evolution of a chaotic system as simulated with different integrators. For each integrator, 500 simulations were run with orbital elements recorded every 1000 years. The orbital elements across all 500 simulations were compiled for each planet into one large pool of data. From this, 2000 points were randomly selected (black dots), so that we were sampling from a range of all simulations. The end result is the distribution at any given time, with increasing density as darker purple. This compares REBOUND's IAS15 with the Mercurius default (Hillfactor = 3), and adjusted to a larger switchover radius (Hillfactor = 7).}
\label{fig:reboundstats}
\end{figure}

 For each integrator, the four giant planets were initialized in the  simulation with randomly selected semimajor axes. Jupiter's semimajor axis was selected from between 5.3 and 6 au, with Saturn then placed in a 3:2 MMR following precedent as in \citet{batygin2010early, masset2001reversing, morbidelli2007dynamics,pierens2008constraints}. We selected a stable multiresonant configuration from \citet{batygin2010early} that was compatible with a Nice model evolution in which Uranus and Neptune continued the MMR chain, with Uranus in 4:3 MMR with Saturn and Neptune in a 4:3 MMR with Neptune. The planets were given eccentricities and inclinations of 0.001, as in \citet{tsiganis2005origin}. Migration and damping forces were added. Approximately 500 simulations were run for each set of integrator parameters, and orbital elements over time were saved for each. From these, 2000 timepoints from all the simulations were randomly selected and plotted on a density distribution as shown in Figure \ref{fig:reboundstats}. 
 For the Mercurius integrator, we tested varying the Hill factor parameter (the radius at which the integrator switches from WHFast to IAS15 to resolve a close encounter) along with the minimum timestep. Our default WHFast timestep was 0.0002 of Jupiter's period, which ranged from 0.015 to 0.018 in simulation time units ($\sim$ 0.002 years) depending on the initial semimajor axis. This is an order of magnitude larger than the minimum orbital time for an encounter, estimated by the period of a body orbiting the at the surface of Jupiter. We also compared our timestep to the crossover time, the time required to traverse the three hill radii crossover radius $r_{c}$. We estimated this as $r_{c}/(e v_{k})$, where the eccentricity $e$ was set to 0.4 and $v_{k}$ is the keplerian velocity, finding $\sim$ 1 year for Jupiter. The minimum IAS timestep for close encounters was set to 0.001 of the default WHFast timestep, which is well within the minimum encounter time and the crossover time. We found that the density distributions for IAS15 and Mercurius were not consistent with each other. Increasing the Hill factor improved the consistency between the two; however, the Hill factor needed negated the purpose of using a hybrid integrator, as the integration would fall entirely in the adaptive timestep regime. Reducing the WHFast timestep did not lead to significant improvement in consistency between the Mercurious and IAS15 outcomes. It appeared that in this chaotic system the Mercurius hybrid integrator was not identifying sufficient close encounters. As a result, we elected to use IAS15 for our integrations.

\section{Additional figures}\label{App:figs}

\begin{figure}[h!]
\includegraphics[trim={0 .45cm 0 .45cm},clip, width=\textwidth]{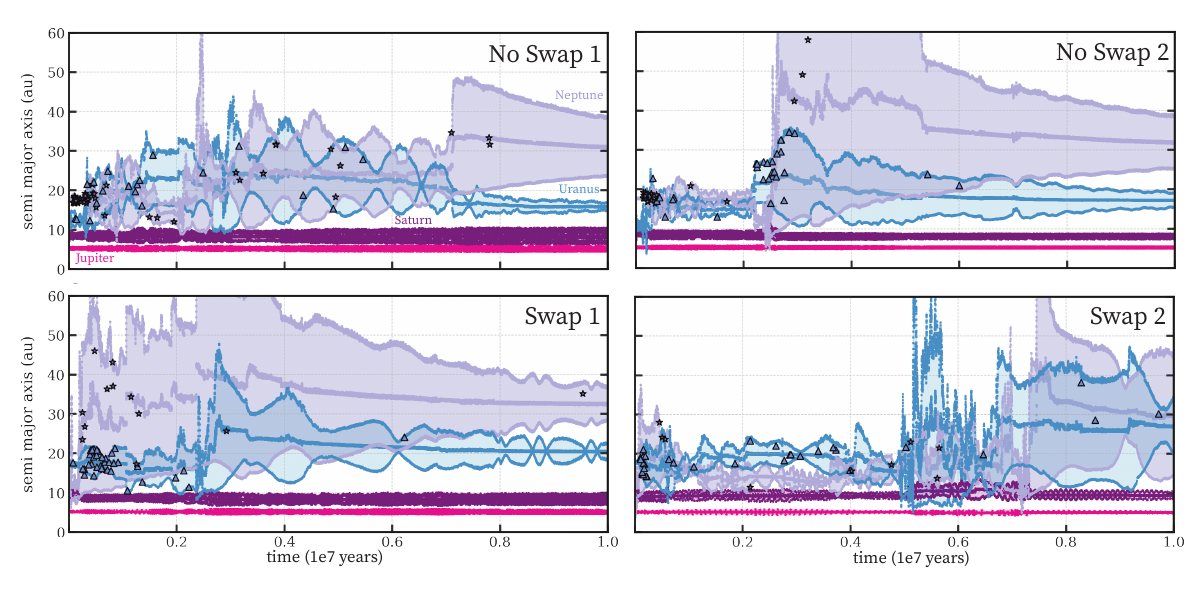} 
\caption{Semimajor axis (a), apocenter distance $a(1+e)$, and pericenter distance $a(1-e)$ vs time of Jupiter (pink), Saturn (dark purple), Uranus (blue), and Neptune (light purple) along with the time and location of collisions with test particles for Uranus (blue triangles) and Neptune (purple stars) during the initial chaotic period. The first row shows simulations in which Neptune starts and ends exterior to Uranus (No Swaps 1\&2) while the second row shows simulations in which Neptune starts interior to Uranus (Swaps 1\&2).}
\label{fig:allrunscollisions}
\end{figure} 

\begin{figure}[h!]
\includegraphics[width=\textwidth]{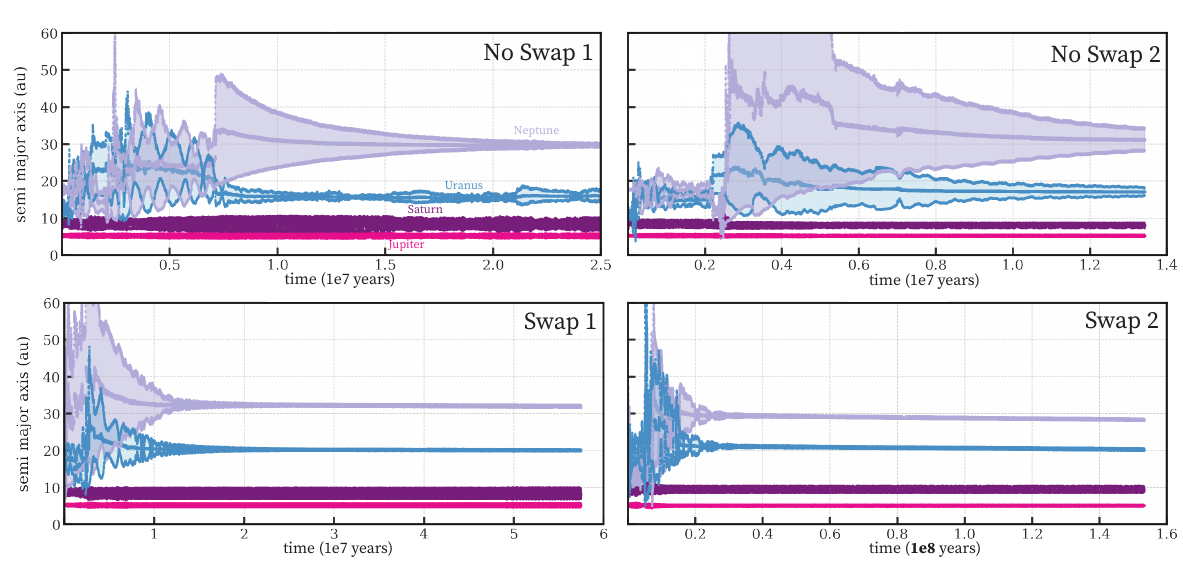} 
\caption{Semimajor axis (a), apocenter distance $a(1+e)$, and pericenter distance $a(1-e)$ vs time of Jupiter (pink), Saturn (dark purple), Uranus (blue), and Neptune (light purple) for the full orbital evolution of the four chosen simulations. The first row shows simulations in which Neptune starts and ends exterior to Uranus (No Swaps 1\&2) while the second row shows simulations in which Neptune starts interior to Uranus (Swaps 1\&2). Note the differing timescales between plots.}
\label{fig:allruns}
\end{figure} 

\end{document}